\documentclass[10pt,a4paper]{article}

\usepackage[margin=2cm]{geometry}
\usepackage{amsmath,amssymb,amsthm}
\numberwithin{equation}{section}
\usepackage{mathtools}
\usepackage{bm}
\usepackage{graphicx}
\usepackage{adjustbox}
\usepackage{xcolor}
\usepackage{booktabs}
\usepackage{array}
\usepackage{multirow}
\usepackage{enumitem}
\usepackage{natbib}
\usepackage{microtype}
\usepackage{fancyhdr}
\usepackage{algorithm}
\usepackage{algpseudocode}
\usepackage{tikz}
\usetikzlibrary{arrows.meta, positioning, fit, backgrounds, calc, shapes.geometric, shadows.blur}
\usepackage[section]{placeins}
\usepackage[affil-it]{authblk}

\usepackage{hyperref}

\definecolor{oiblue}{HTML}{2878B5}
\definecolor{oigreen}{HTML}{299D70}
\definecolor{oiverm}{HTML}{A64D79}
\definecolor{ink}{HTML}{263238}
\definecolor{mut}{HTML}{78909C}

\theoremstyle{plain}
\newtheorem{theorem}{Theorem}[section]
\newtheorem{proposition}[theorem]{Proposition}
\newtheorem{lemma}[theorem]{Lemma}

\theoremstyle{definition}
\newtheorem{definition}[theorem]{Definition}

\newtheorem{remark}[theorem]{Remark}

\DeclareMathOperator*{\argmin}{arg\,min}

\newcommand{\R}{\mathbb{R}}
\newcommand{\E}{\mathbb{E}}

\newcommand{\Prob}{\mathbb{P}}
\newcommand{\Q}{\mathbb{Q}}
\newcommand{\calL}{\mathcal{L}}
\newcommand{\calN}{\mathcal{N}}
\newcommand{\calF}{\mathcal{F}}
\newcommand{\calM}{\mathcal{M}}

\newcommand{\Jop}{\mathcal{J}}       %
\newcommand{\Vb}{\overline{V}}
\newcommand{\norm}[1]{\left\|#1\right\|}
\newcommand{\abs}[1]{\left|#1\right|}
\newcommand{\PIRanNN}{\text{PIRanNN}}
\newcommand{\reg}{\gamma}            %
\newcommand{\vbar}{\bar v}           %
\newcommand{\nens}{n_{\mathrm{ens}}} %
\newcommand{\nfeat}{m}               %
\newcommand{\npar}{n_\theta}         %

\newcommand{\chg}[1]{{#1}}
\newcommand{\chgb}[1]{{#1}}
\definecolor{chggreen}{HTML}{1B7F3B}
\newcommand{\chgg}[1]{{#1}}
\definecolor{chgorange}{HTML}{D55E00}
\newcommand{\chgo}[1]{{#1}}

\hypersetup{
  colorlinks=true,
  linkcolor=blue!70!black,
  citecolor=green!50!black,
  urlcolor=blue!70!black
}

\title{
  \vspace{-1.5cm}
  \textbf{PIRanNN: Random-Feature Pricing and Calibration\\
  under Jump-Diffusion and Stochastic-Volatility Models}
}

\author[1,2]{Saurabh Bansal\thanks{Corresponding author:
  \texttt{sbansal@rgipt.ac.in}}}
\author[3]{Srikanth K. Iyer}
\author[2]{Shashi Jain}

\affil[1]{Department of Mathematical Sciences, Rajiv Gandhi Institute of
  Petroleum Technology, Jais, Amethi 229304, Uttar Pradesh, India}
\affil[2]{Department of Management Studies, Indian Institute of Science,
  Bengaluru 560012, Karnataka, India}
\affil[3]{Department of Mathematics, Indian Institute of Science,
  Bengaluru 560012, Karnataka, India}

\date{\empty}

\begin{document}
\maketitle
\thispagestyle{fancy}

\begin{abstract}
\chgg{We study a random-feature method for pricing and calibrating European
options under jump-diffusion (Merton, Kou) and stochastic-volatility
(Heston, SABR) models. The network has one hidden layer with a probit
activation and fixed random inner weights, so fitting it to the pricing
equation, the payoff and the boundary conditions is a single regularised
least-squares problem, with no iterative training. For Merton and Kou the
jump integral of each feature is explicit, so no quadrature is needed. One
factorisation of the least-squares matrix gives the price at every state
and time, its Greeks, and the exact derivatives of the prices with respect
to the model parameters for gradient-based calibration. For the jump
models, subtracting the Black--Scholes price removes the payoff kink and
reduces the pricing error by about two orders of magnitude. The method is
then about as accurate as a standard Crank--Nicolson finite-difference
scheme at the same run time, although a finite-difference scheme that
also subtracts the Black--Scholes price is about two orders of magnitude
more accurate still. On the test problems of a recent physics-informed
neural network study, the method matches that network's accuracy in
seconds instead of about half an hour. The calibration experiments show
that a solver can fit the quotes and still return wrong parameters when
its pricing error is not small compared with the effect of the parameters
on the prices: without the subtraction the solver misses the jump
parameters by up to $50\%$, and with it they are recovered to within
$3\%$. For
Heston and SABR, used as an applicability study with boundary conditions
that need no reference prices,} \chgo{pricing errors are about one
percent} \chgg{and the calibrated parameters are within about $15\%$
of the true values.}
\end{abstract}


\bigskip\hrule

\section{Introduction}\label{sec:intro}

\subsection{The problem}\label{sec:intro-problem}

The Black--Scholes model prices a European option in closed form, but it
does not reproduce what is observed in option markets. Implied volatility
varies with strike and maturity, asset prices occasionally jump, and
volatility itself fluctuates over time. Two families of models address
these features. Jump-diffusion models add random jumps to the asset
price; we consider the models of \citet{merton1976}, with normally
distributed log-jumps, and of \citet{kou2002}, with double-exponential
log-jumps. Stochastic-volatility models let the variance follow its own
diffusion, correlated with the asset price; we consider the models of
\citet{heston1993} and \citet{hagan2002} (SABR).

In each of these models the price of a European option is the expected
discounted payoff under the risk-neutral measure. By the Feynman--Kac
theorem, the price as a function of the current state and the time to
maturity solves a linear partial differential equation (PDE). For the
jump models it solves a partial integro-differential equation (PIDE): a
PDE with an additional integral term that accounts for the jumps.
\chg{For the four models above, European prices can also be computed
quickly and accurately by Fourier methods
\citep{carr1999,fang2008}. We use these models because such accurate
references exist, which lets us measure the error of the method. The PDE
route is of most interest for products and settings where transform
methods are less convenient, for example continuously monitored barrier
options under stochastic volatility, or the repeated valuation of a
portfolio along simulated paths.}

In practice, pricing goes hand in hand with calibration. A model is first
calibrated: its parameters are chosen so that the model reproduces the
quoted prices of liquid options, typically vanilla calls and puts. The
calibrated model is then used to price and hedge less liquid or
non-standard derivatives on the same asset. Calibration requires pricing
the quoted options at many trial parameter vectors. Gradient-based
optimisers, such as Gauss--Newton and Levenberg--Marquardt methods, also
need at each
trial vector the derivative of every quoted price with respect to every
parameter. A method used for both tasks should therefore deliver prices
and these derivatives cheaply and reliably.

\subsection{Neural-network solvers and random features}\label{sec:intro-nn}

\chg{Why use neural networks for pricing equations at all? For European
options in one or two state variables, Fourier methods and finite
differences are fast, accurate and supported by convergence theory, and
they are hard to improve upon. The case for neural networks rests on
problems that lie outside this setting. The most cited is a large number
of state variables, for which grids become infeasible; this motivated
mesh-free methods such as the deep Galerkin method \citep{sirignano2018}
and the deep BSDE method \citep{han2018}. A second is the repeated
evaluation of one pricing map, for example during calibration, where a
network trained once can then be evaluated almost instantly
\citep{horvath2021,hainaut2024}. But even in low dimension there are two
common situations in which classical methods are less convenient.
Non-standard payoffs, such as barrier or Asian options under jump or
stochastic-volatility models, usually have no Fourier solution and need a
problem-specific grid. And exposure and margin calculations need prices
and sensitivities of a trade at many states and future dates along
simulated paths, which calls for a price surface that can be evaluated
and differentiated anywhere. A neural network provides exactly such a
surface, defined at every state and time and differentiable everywhere.
These advantages come at a price: training is a slow, nonconvex
optimisation whose result depends on the initialisation, and it offers
no error control comparable to refining a grid.} \chgg{Systematic
comparisons bear this out. On standard test equations in one to three
space dimensions, PINNs have not matched finite elements in accuracy or
solution time \citep{grossmann2024}, and a review of machine-learning
solvers for partial differential equations found that most claims of
outperforming classical methods rested on comparisons with weak
baselines \citep{mcgreivy2024}.}

Physics-informed neural networks (PINNs) \citep{raissi2019} solve a PDE by
training a neural network so that it nearly satisfies the equation, the
initial condition and the boundary conditions at a large set of sample
points. They are flexible, but training is a nonconvex optimisation: it
is slow, it can stop at different solutions depending on the random
initialisation, and each step needs derivatives of the network output with
respect to its inputs, usually obtained by automatic differentiation. For
jump models the integral term must in addition be approximated by a
quadrature rule at every sample point \citep{bansal2026}.

Random-feature methods remove the training loop. The inner weights of a
one-hidden-layer network are drawn at random and then held fixed, so the
network output is a linear combination of fixed functions (the
\emph{features}) with unknown coefficients (the \emph{output weights}).
The pricing equation is linear: each of its terms is a known coefficient
multiplied by the unknown price function, by one of its derivatives, or,
for the jump models, by its average over a jump. If the network output is
substituted for the price, the equation at any given point therefore
becomes a linear equation in the output weights. Imposing it at many
sample points, together with the payoff and the boundary conditions, gives
a system of linear equations for the output weights, which is solved once
by least squares. This idea, known as
the extreme learning machine, has been applied to PDEs from physics
\citep{dwivedi2020,dong2021,chen2022rfm,wang2023} and to option pricing
\citep{gonon2023,srinivasan2025}.

\subsection{Approach and contributions}\label{sec:contributions}

\chgb{We follow this random-feature approach, with features that apply the
standard normal distribution function $\Phi$ (the probit function) to a
random linear combination of the state and the time to maturity, and we
call the method \PIRanNN{} (physics-informed random neural network). Fitting
random features to a linear pricing equation by least squares is not new
in itself: it is the physics-informed extreme learning machine of
\citet{dwivedi2020}, which has already been applied to the Black--Scholes
and Heston--Hull--White equations \citep{srinivasan2025}
(Section~\ref{sec:related-rf}). Our contributions are the following.}

\begin{enumerate}[leftmargin=*]
\item \chg{\chgb{\emph{The jump term is exact.}} We use the
  standard normal distribution function $\Phi$ as activation. Averaging a
  probit feature over a normal or a double-exponential jump gives an
  explicit formula (Lemma~\ref{lem:jump}). Every term of the pricing
  equation, including the jump integral of the Merton and Kou models, can
  therefore be evaluated exactly for each feature, and the resulting PIDE
  solver needs no quadrature.}
\item \chgb{\emph{One factorisation serves many purposes.} The matrix of the
  least-squares problem is factorised once, for the price, and the
  factorisation is then reused. The Greeks are derivatives of the fitted
  price with respect to the state, available at every point of the domain.
  Differentiating the least-squares solution with respect to the model
  parameters gives the exact Jacobian of the solver's prices, for
  diffusion and jump parameters alike (Proposition~\ref{prop:sens}); each
  parameter costs only a small fraction of the first solve, so calibration
  needs no offline training over a parameter range. The current variance
  $v_0$ of Heston and the current volatility $\alpha_0$ of SABR only move
  the point at which the price surface is evaluated, so recalibrating them
  needs no new solve. Other European payoffs under the same model can also
  be priced with the same factorisation (Section~\ref{sec:lsq}). These
  properties follow from linearity, and the Jacobian formula is the
  classical one of \citet{golub1973}; what we add is their combination in
  one tool for pricing and calibration.} \chgg{Because the fitted price is a
  formula in the state and the time to maturity, all of this is available
  at any point, with no interpolation on a grid.}
\item \chg{\chgb{\emph{For the jump models, \chgo{a quadrature-free Black--Scholes subtraction} removes the main source of error.}} The call
  payoff has a kink at the strike, which a sum of smooth features
  represents poorly at short maturities. We subtract the Black--Scholes
  price, which already contains the kink, and solve for the remainder\chgo{,
  a classical technique (Section~\ref{sec:bs-subtraction})}. The
  jump integral of the Black--Scholes price is also explicit, so this step
  remains quadrature-free.} \chgg{It is the main source of accuracy on the
  jump models: in the ablation of Section~\ref{sec:merton-exp} it reduces
  the error by a factor of about $100$ for Merton and $70$ for Kou, and by
  about $700$ and $300$ together with extra features near the strike, for
  the same total number of features. The step is not specific to random
  features; it improves a finite-difference scheme by a similar factor.}
  \chgo{For Heston and SABR, the price of the model with the volatility of
  volatility set to zero, a classical approximation, can be subtracted in
  the same way; in our tests it reduces the error for both models, by an
  amount that varies from case to case
  (Sections~\ref{sec:heston-exp} and~\ref{sec:sabr-exp}).}
\item \chg{\emph{Solver accuracy limits identifiability.}} \chgg{In the
  calibration of both jump models, the plain solve fits the quoted prices
  but returns jump parameters that are off by up to $50\%$, because its
  pricing error is larger than the change in the prices caused by the
  parameter error. If the calibration is repeated with a new random draw
  of the inner weights, it again fits the quotes but returns different,
  equally wrong, jump parameters. With the
  Black--Scholes subtraction the pricing error is $30$ to $50$ times
  smaller and the parameters are recovered to within $3\%$
  (Section~\ref{sec:jump-calib}).} \chg{This differs from the known intrinsic
  difficulty of separating jump and diffusion parameters, where the
  quoted prices themselves do not determine the parameters whatever the
  pricer \citep{albani2020}: here the parameters are identifiable, and
  the loss is caused by the solver. Indeed, when the quotes are generated
  by the plain solver itself\chgg{, with the same draw of the inner
  weights}, the same solver recovers the parameters
  exactly. The same consideration applies to any
  calibration that uses an approximate pricer.} \chgg{The exact Jacobian,
  and repeating the calibration with new random draws of the inner
  weights, let one check a calibration for this problem without a
  reference price
  (Section~\ref{sec:identifiability}).}
\end{enumerate}

\chgb{We test the method on European options in
models with one or two state variables, where accurate references exist.
The jump models receive the full treatment above.} \chgg{There we compare
it with a Crank--Nicolson finite-difference scheme with an implicit jump
term, at equal run time and with the same boundary data, which is the
kind of strong baseline that \citet{mcgreivy2024} call for. With the
Black--Scholes subtraction, our method is about as accurate as this
scheme: more accurate for run times of a few seconds and less accurate
for longer ones, because its error levels off near $10^{-6}$ (relative to
the strike) while the finite-difference error keeps falling. A
finite-difference scheme that also subtracts the Black--Scholes price is
about two orders of magnitude more accurate than either. We therefore do
not claim that the method is the most accurate pricer in one space
dimension. Its case rests on what one solve provides: a price that is a
formula in the state and time, its Greeks and its exact parameter
Jacobian from one factorisation, a jump term without quadrature, and a
formulation that carries over unchanged to two state variables.}
\chgb{Heston and SABR are an applicability
study: we use only simple boundary conditions that need no reference
prices} \chgo{and, for the main results, the plain solve, which is what one
has when no suitable function to subtract is known.} \chgb{The method then
prices and calibrates equations whose coefficients depend on the state,
at an accuracy of about one percent, \chgo{partly} limited by the payoff kink; for
Heston, the best ADI schemes remain more accurate in three of the four
standard test cases.}

\chgb{The method is also transparent. For a given draw of the features, the output
weights are the unique solution of a regularised least-squares problem,
whereas a trained PINN depends on its random initialisation, the
optimiser and the point at which training is stopped. The fitted price,
its Greeks and its Jacobian are explicit formulas in known functions, and
this is what makes the checks of the fourth item possible.}

\chgb{The method keeps the mesh-free
formulation and the smooth price surface of neural-network solvers but
removes the training step, so the applications of
Section~\ref{sec:intro-nn} are direct extensions. A continuously
monitored barrier adds a boundary condition at the barrier, and for the
Merton model the jump term then remains explicit
(Section~\ref{sec:conclusion}). An Asian option adds the running average
as a state variable, and Lemma~\ref{lem:pointwise} applies unchanged to an
operator with more state variables. For exposure and margin calculations,
one solve gives prices and sensitivities at every state and time in the
domain, so valuing a trade along simulated paths needs no further solves.
These extensions are not tested here.}

\subsection{Outline}\label{sec:outline}

Section~\ref{sec:related} reviews related work, including the different
ways in which neural-network pricers have been used for calibration.
Section~\ref{sec:models} sets out the models, their pricing equations and
the boundary conditions. Section~\ref{sec:method} describes the method:
the features, the least-squares problem, the explicit formulas (first for
Merton, then in general), the choice of loss weights and the
Black--Scholes subtraction. Section~\ref{sec:calibration} derives the
Greeks, the parameter sensitivities and the calibration procedure.
Section~\ref{sec:experiments} reports the numerical experiments, and
Section~\ref{sec:conclusion} concludes.

\section{Related work}\label{sec:related}

\subsection{Classical numerical methods}\label{sec:related-classical}

For European options under jump-diffusion and affine stochastic-volatility
models, the most efficient methods use the characteristic function of the
log-price. The fast Fourier transform method of \citet{carr1999} and the
Fourier-cosine (COS) method of \citet{fang2008} compute a European price
from a short series whose terms involve the characteristic function and
the Fourier coefficients of the payoff. When no characteristic function is
available, or when the product is not European, the pricing equation is
usually solved by finite differences. For jump models the integral term is
then approximated by quadrature at every time step \citep{cont2005}. For
two-dimensional stochastic-volatility equations, alternating-direction
implicit (ADI) schemes are the standard choice \citep{inthout2010}, and
the BENCHOP-SLV project \citep{vonsydow2019} provides highly accurate
reference prices for the Heston model.

\chg{Calibration with a finite-difference pricer is also well studied. The
gradient of the prices with respect to the model parameters can be
computed by differentiating the discretised equation, which leads to
adjoint methods; \citet{albani2020} use such a gradient to calibrate
jump-diffusion models with local volatility. Differentiating the solution
of a least-squares problem with respect to parameters on which its matrix
depends is also classical \citep{golub1973}. An exact parameter gradient
of a PDE pricer is therefore not new in itself. What is new here is
obtaining it for a mesh-free, quadrature-free solver at
\chgb{a small fraction of the cost of the price itself}.}

\subsection{Physics-informed neural networks for pricing and calibration}\label{sec:related-pinn}

PINNs \citep{raissi2019} have been extended in many directions, for
example to periodic activation functions, which have also been used to fit
PDE residuals \citep{sitzmann2020}. In option pricing, \citet{bansal2026} apply PINNs to the
Merton and Kou PIDEs, approximating the jump integral by the composite
trapezoidal rule, and compare with finite differences; \citet{bansal2025}
extend the approach to weather derivatives. For stochastic volatility,
\citet{hainaut2024} train a PINN on the Heston pricing equation over a
range of option specifications and model parameters, so that new prices
require no retraining. \citet{mrad2025} combine supervised learning on
simulated payoffs with a PDE penalty for Heston with time-dependent
parameters. \citet{bae2024} solve the Black--Scholes equation with a given
local volatility surface by a PINN, and check the result by recovering
the surface from the trained network through Dupire's formula.

\chg{Physics-informed networks have also been used for calibration, in two
main ways.}
\begin{itemize}[leftmargin=*]
\item \chg{\emph{Inverse PINNs} treat the unknown model parameters as
  additional trainable variables and fit them together with the network
  weights, as in the parameter-identification examples of
  \citet{raissi2019}. In finance, this approach has been used to calibrate
  the market price of risk in a jump-diffusion PIDE for catastrophe bonds,
  from synthetic prices \citep{ipinn2026}. The network and the
  parameters are found by a single nonconvex training run, which must be
  repeated whenever the market data change.}
\item \chg{\emph{Parametric PINNs} and physics-informed operator networks
  take the model parameters as additional inputs and are trained once,
  offline, over a box of parameter values. After training, a price at any
  parameter vector in the box is a fast evaluation, and its derivative
  with respect to the parameters is available by automatic
  differentiation. \citet{hainaut2024} take this approach for Heston, and
  \citet{chakhoyan2025} use a physics-informed DeepONet with the aim of
  real-time Heston calibration. The accuracy at a given parameter vector
  is fixed by the offline training and cannot be improved at that vector
  without retraining.}
\end{itemize}

\chg{The present method takes a third route. At each trial parameter
vector the pricing problem is solved directly, by a single convex
least-squares problem, and its exact derivative with respect to the
parameters is obtained from the same factorisation.
Table~\ref{tab:calib-routes} summarises the differences.}

\begin{table}[htbp]
\centering
\caption{\chg{Three ways of calibrating with a physics-informed
network.}}
\label{tab:calib-routes}
\small
\begin{tabular}{>{\raggedright}p{0.17\textwidth}>{\raggedright}p{0.23\textwidth}>{\raggedright}p{0.23\textwidth}>{\raggedright\arraybackslash}p{0.27\textwidth}}
\toprule
 & Inverse PINN & Parametric PINN / operator network & \PIRanNN{} (this paper) \\
\midrule
What is fitted & network weights and model parameters together & network weights, offline, over a parameter box & output weights, at each trial parameter vector \\
Offline cost & none & large & none \\
Price at a trial parameter vector & not computed separately & evaluation of the trained surrogate & one regularised least-squares solve \\
Derivative in the parameters & implicit in the joint training & automatic differentiation of the surrogate & \leavevmode\chgb{exact; reuses the factorisation of the price} \\
Parameter range & unrestricted & limited to the training box & unrestricted \\
Fitting the network & nonconvex & nonconvex (offline) & convex (least squares) \\
\bottomrule
\end{tabular}
\end{table}

\subsection{Random-feature methods}\label{sec:related-rf}

Random-feature methods sit between deep networks and classical spectral
methods. A single hidden layer of features is drawn from a fixed
distribution and held fixed, and only the output weights are fitted.
\citet{rahimi2008} show that if the target function is an average of the
features over their sampling distribution, with a weight bounded relative
to that distribution, then with high probability a combination of $\nfeat$
randomly drawn features approximates it with an $L^2$ error of order
$\nfeat^{-1/2}$.
\citet{dwivedi2020} introduce the physics-informed extreme learning
machine, in which the output weights are fitted to the PDE residual by
least squares. \citet{dong2021} combine extreme learning machines with
domain decomposition for linear and nonlinear PDEs, \citet{chen2022rfm}
develop a closely related random feature method, and \citet{wang2023}
extend the approach to dimensions up to thirty. Like the present method, these works build a
random-feature approximation, impose the equation at random sample points
and solve a linear least-squares problem. \chg{Computing the derivatives
of a feature explicitly is straightforward for any smooth activation
function, including the tanh activation used in these works; it is not
what distinguishes the present method. What does is the explicit jump
integral, which does not arise for the equations they consider, and the
derivative with respect to the model parameters.}

Closest to our setting, \citet{srinivasan2025} apply an extreme learning
machine to the Black--Scholes and Heston--Hull--White equations and
include a calibration experiment based on Bayesian optimisation.
\chg{Bayesian optimisation does not use derivatives, so the linear system
is solved afresh at every trial parameter vector. The exact Jacobian of
Proposition~\ref{prop:sens} allows a derivative-based optimiser instead.}

\citet{gonon2023} proves error bounds without the curse of
dimensionality for random-feature networks that learn option prices in
Black--Scholes and exponential L\'evy models, including models with
jumps. There the network is fitted by regression to samples whose
conditional mean is the price, such as simulated payoffs, rather than to
the pricing equation itself.
\citet{neufeldschmocker2023}, \citet{neufeld2025} and \citet{beck2021}
develop random-feature and deep-splitting methods for semilinear PIDEs,
using time stepping and regression of conditional expectations.
\citet{jacquier2023} use random neural networks for path-dependent options
under rough volatility. The Laguerre network of \citet{chen2021} also uses
extreme-learning-machine training, but with a fixed polynomial basis:
polynomial bases converge very fast for smooth targets, but their jump
integrals are not available in closed form.

\section{Models and pricing equations}\label{sec:models}

This section sets out the four models and the equations their prices
satisfy. All four equations have the same structure, which the method of
Section~\ref{sec:method} exploits: they are linear in the unknown price
function, and their coefficients are explicit functions of the state.

\subsection{Setting}\label{sec:setting}

We follow the standard arbitrage-free pricing framework, as presented for
diffusion models in \citet{shreve2004} and for jump models in
\citet{cont2004book}. Let $S_t$ be the price at time $t$ of the
underlying asset, which pays a continuous dividend yield $q\ge0$ (for a
currency, $q$ is the foreign interest rate). We consider a European
derivative on this asset that pays $g(S_T)$ at maturity $T$, for a given
payoff function $g$, and denote by $V(S,t)$ its price at time $t$ when
$S_t=S$.

For ease of notation, we take the interest rate $r$ to be constant, so the
bank account is $B_t=e^{rt}$. Prices are computed under a risk-neutral
measure $\Q$: a probability measure, equivalent to the real-world measure,
under which the value of every traded asset, with any dividends
reinvested, divided by $B_t$ is a martingale. For the underlying, this means
that $e^{-(r-q)t}S_t$ is a martingale, and the price of the derivative is
the expected discounted payoff,
\begin{equation}\label{eq:price-expectation}
V(S,t) = \E_\Q\big[\,e^{-r(T-t)}g(S_T) \,\big|\, S_t=S\,\big].
\end{equation}
For the jump and stochastic-volatility models considered here, the
risk-neutral measure is not unique, and as usual we specify each model
directly under $\Q$. In the stochastic-volatility models the state
contains, besides $S_t$, the current variance (Heston) or volatility
(SABR), and the conditioning in \eqref{eq:price-expectation} is on the
whole state. SABR is set under the $T$-forward measure instead
(Section~\ref{sec:sabr-model}).

The least-squares method of Section~\ref{sec:method} applies to any such
payoff: $g$ enters only through the condition at maturity and the boundary
data. For definiteness we use the European call, $g(S)=(S-K)^+=\max(S-K,0)$ with strike
$K$, as the running example, and all experiments with the four models
price calls. \chg{Other products change the problem in identifiable ways:
a continuously monitored barrier adds a boundary condition at the barrier,
and an Asian option adds the running average of the asset price as a
further state variable. We do not treat them here.}

For the call, to obtain a single payoff for all strikes, we use the log-moneyness
$x=\log(S/K)$ and the time to maturity $\tau=T-t$, and we measure prices in
units of the strike:
\begin{equation*}
\Vb(x,\tau) = \frac{1}{K}\,V\big(Ke^x,\,T-\tau\big).
\end{equation*}
At maturity ($\tau=0$) the normalised price equals the normalised payoff
\begin{equation*}
\Vb(x,0) = f(x), \qquad f(x) = (e^x-1)^+ .
\end{equation*}

Let $X_t=\log(S_t/K)$ be the log-moneyness process (together with the
variance or volatility, for the stochastic-volatility models). The models
below are time-homogeneous, so \eqref{eq:price-expectation} becomes, in
these variables,
\[
\Vb(x,\tau) = \E_\Q\big[\,e^{-r\tau} f(X_\tau) \,\big|\, X_0 = x\,\big].
\]
By the Feynman--Kac theorem, this expectation solves the equation
\begin{equation}\label{eq:fk}
\partial_\tau \Vb = \calL \Vb - r\Vb,
\end{equation}
where $\calL$ is the generator of $X$, defined for smooth $u$ by
$(\calL u)(x)=\displaystyle \lim_{h\to0^+}\big(\E[u(X_h)\mid X_0=x]-u(x)\big)/h$.
For each model below we give $\calL$ explicitly. It is convenient to
collect all terms on one side and write the equation as
\begin{equation}\label{eq:M-generic}
\calM[\Vb] = 0, \qquad \calM[u] := \partial_\tau u - \calL u + r u .
\end{equation}
We call $\calM$ the \emph{pricing operator}. It is linear in $u$:
$\calM[au+bw] = a\,\calM[u]+b\,\calM[w]$ for constants $a,b$ and functions
$u,w$. This linearity is what makes the method of
Section~\ref{sec:method} a linear least-squares problem.

\subsection{Jump-diffusion models: Merton and Kou}\label{sec:pde}

In an exponential L\'evy jump-diffusion model the asset price follows
\begin{equation*}
\frac{dS_t}{S_{t^-}} = (r - q - \lambda\kappa_J)\,dt + \sigma\,dW_t + (e^{J} - 1)\,dN_t,
\end{equation*}
where $W$ is a standard Brownian motion, $N$ is a Poisson process with
intensity $\lambda\ge0$, and each jump multiplies the price by $e^J$,
where the log-jump $J$ has density $g$ and is independent of $W$ and $N$.
The constant $\kappa_J=\E[e^J-1]$ is the mean relative jump size; the
term $-\lambda\kappa_J$ in the drift compensates for the jumps, so that
the discounted price is a martingale.

The generator of the log-moneyness has a diffusion part and a jump part.
The jump part is the jump rate $\lambda$ times the expected change in $u$
caused by one jump. We write the expected value of $u$ just after a jump as
\begin{equation*}
(\Jop u)(x,\tau) := \E\big[u(x+J,\tau)\big] = \int_\R u(x+y,\tau)\,g(y)\,dy .
\end{equation*}
The generator is then
\[
\calL u = \tfrac12\sigma^2\,\partial_{xx}u + \mu_0\,\partial_x u
        + \lambda\big(\Jop u - u\big),
\qquad
\mu_0 = r - q - \tfrac12\sigma^2 - \lambda\kappa_J ,
\]
and the pricing equation \eqref{eq:fk} becomes the PIDE
\begin{equation*}
\partial_\tau \Vb = \tfrac{1}{2}\sigma^2 \partial_{xx} \Vb
  + \mu_0 \, \partial_x \Vb - (r + \lambda)\Vb
  + \lambda \,\Jop\Vb .
\end{equation*}
In the form \eqref{eq:M-generic}, the pricing operator is
\begin{equation}\label{eq:M-jump}
\calM[u]=\partial_\tau u-\tfrac12\sigma^2\partial_{xx}u-\mu_0\,\partial_xu
+(r+\lambda)\,u-\lambda\,\Jop u .
\end{equation}
The first four terms are local: they involve $u$ and its derivatives at
the point $x$ only. The last term is non-local: it involves the values of
$u$ at all points $x+y$ that a jump can reach.

\chgb{In the Merton model \citep{merton1976}, the} log-jump is normal,
$J\sim\calN(\mu_J,\sigma_J^2)$:
\begin{equation*}
g(y) = \frac{1}{\sqrt{2\pi}\,\sigma_J}
  \exp\!\left(-\frac{(y - \mu_J)^2}{2\sigma_J^2}\right),
\qquad
\kappa_{J} = e^{\mu_J + \frac{1}{2}\sigma_J^2} - 1.
\end{equation*}

\chgb{In the Kou model \citep{kou2002}, the} log-jump is upward with
probability $p$ and then exponentially distributed with rate $\eta_1$, and
downward with probability $1-p$ and then exponentially distributed with
rate $\eta_2$:
\begin{equation}\label{eq:g-kou}
g(y) = p\eta_1 e^{-\eta_1 y}\mathbf{1}_{\{y \ge 0\}}
  + (1-p)\eta_2 e^{\eta_2 y}\mathbf{1}_{\{y < 0\}},
\qquad
\kappa_{J} = \frac{p\eta_1}{\eta_1 - 1}
  + \frac{(1-p)\eta_2}{\eta_2 + 1} - 1 .
\end{equation}
The condition $\eta_1>1$ ensures that $\E[e^J]$, and hence $\kappa_J$, is
finite. The double-exponential law has heavier tails than the normal law,
and its upward and downward tails can differ.

\subsection{Heston model}\label{sec:heston-model}

In the model of \citet{heston1993} the variance $v$ of the asset follows
its own mean-reverting diffusion:
\begin{equation*}
dS_t = (r-q) S_t\,dt + \sqrt{v_t}\, S_t\, dW_t^{(1)},
\qquad
dv_t = \kappa(\vbar - v_t)\,dt + \xi\sqrt{v_t}\,dW_t^{(2)},
\end{equation*}
where the Brownian motions have correlation $\rho$. The variance reverts
to the long-run level $\vbar$ at speed $\kappa$, and $\xi$ is the
volatility of the variance. The variance stays strictly positive if the
Feller condition $2\kappa\vbar\ge\xi^2$ holds. The state is now
$(x,v)$, and the pricing equation is the two-dimensional PDE
\begin{equation}\label{eq:heston-pde}
\partial_\tau \Vb = \tfrac{1}{2}v\,\partial_{xx}\Vb
  + \rho\xi v\,\partial_{xv}\Vb
  + \tfrac{1}{2}\xi^2 v\,\partial_{vv}\Vb
  + \big(r - q - \tfrac{v}{2}\big)\partial_x \Vb
  + \kappa(\vbar - v)\partial_v \Vb - r\Vb,
\end{equation}
with pricing operator
\begin{equation}\label{eq:M-heston}
\calM[u]=\partial_\tau u-\tfrac12 v\,\partial_{xx}u-\rho\xi v\,\partial_{xv}u
-\tfrac12\xi^2v\,\partial_{vv}u-\big(r-q-\tfrac v2\big)\partial_xu
-\kappa(\vbar-v)\,\partial_vu+r\,u .
\end{equation}
There is no jump term, but the coefficients now depend on the state
through $v$. Two of the benchmark cases of Section~\ref{sec:heston-exp}
are currency options, with $q>0$ the foreign interest rate.

\subsection{SABR model}\label{sec:sabr-model}

The SABR model of \citet{hagan2002} describes a forward price $F$ and its
volatility $\alpha$:
\begin{equation}\label{eq:sabr-sde}
dF_t = \alpha_t\, F_t^\beta\, dW_t^{(1)},
\qquad
d\alpha_t = \nu\,\alpha_t\, dW_t^{(2)},
\end{equation}
with correlation $\rho$ between the two Brownian motions, exponent
$\beta\in[0,1]$ and volatility of volatility $\nu>0$. Under the
$T$-forward measure the forward is a martingale and prices are expressed
in units of the zero-coupon bond, so there is neither a drift nor a
discounting term. The undiscounted call price $C(F,\alpha,\tau)$ solves
\begin{equation*}
\partial_\tau C = \tfrac{1}{2}\alpha^2 F^{2\beta}\partial_{FF} C
 + \rho\nu\alpha^{2} F^\beta \partial_{F\alpha} C
  + \tfrac{1}{2}\nu^2\alpha^2 \partial_{\alpha\alpha} C ,
\end{equation*}
with pricing operator
\begin{equation}\label{eq:M-sabr}
\calM[u]=\partial_\tau u-\tfrac12\alpha^2F^{2\beta}\,\partial_{FF}u
 -\rho\nu\alpha^{2} F^{\beta}\,\partial_{F\alpha}u
-\tfrac12\nu^2\alpha^2\,\partial_{\alpha\alpha}u .
\end{equation}
Apart from the time derivative, the equation contains only second
derivatives, with coefficients that
depend on the state through $\alpha^2F^{2\beta}$, $\alpha^2 F^\beta$ and
$\alpha^2$. \citet{hagan2002} give an approximate formula for the implied
volatility, which is accurate near the money and for moderate maturities
but degrades for extreme strikes and long maturities. \chg{Because it is
an approximation, it is not an exact reference;
Section~\ref{sec:sabr-exp} also compares with Monte Carlo simulation of
\eqref{eq:sabr-sde}.}

\subsection{Common structure}\label{sec:common}

All four pricing operators, \eqref{eq:M-jump}, \eqref{eq:M-heston} and
\eqref{eq:M-sabr}, are linear in the unknown price function
$u$ and have the form
\begin{equation}\label{eq:M-split}
\calM[u] = \underbrace{\sum_{|\iota|\le 2} a_\iota(z)\,\partial^\iota u + c(z)\,u}_{\text{local part}}
  \;\underbrace{-\;\lambda\,\Jop u}_{\text{jump part}},
\end{equation}
where $z$ is the vector of state variables and time, $\partial^\iota$
ranges over the partial derivatives of order at most two (including
$\partial_\tau$; $\iota$ is a multi-index), and the jump part is present
only for Merton and Kou. The coefficients $a_\iota(z)$ and $c(z)$ are constants for the jump
models and simple explicit functions of the state for Heston and SABR.
The method of Section~\ref{sec:method} needs only two properties of
\eqref{eq:M-split}: that it is linear in $u$, and that its coefficients can
be evaluated at any given point. \chg{The Heston--Hull--White model, with
a stochastic short rate as a third state variable, has the same form\chgb{;
we do not treat it here}.}

\subsection{Notation}\label{sec:notation}

Table~\ref{tab:notation} collects the main symbols.

\begin{table}[htbp]
\centering
\caption{Main notation.}
\label{tab:notation}
\small
\begin{tabular}{>{\raggedright}p{0.17\textwidth}>{\raggedright}p{0.29\textwidth}>{\raggedright}p{0.13\textwidth}>{\raggedright\arraybackslash}p{0.31\textwidth}}
\toprule
Symbol & Meaning & Symbol & Meaning \\
\midrule
$x=\log(S/K)$ & log-moneyness & $\Phi,\ \phi$ & standard normal distribution function and density \\
$\tau=T-t$ & time to maturity & $\nfeat$ & number of random features \\
$\Vb(x,\tau)$ & price in units of the strike & $A_k,\ \zeta_k$ & inner weight vector and centre of feature $k$ \\
$f(x)=(e^x-1)^+$ & normalised call payoff & $s_k(z)$ & pre-activation $A_k\cdot(z-\zeta_k)$ \\
$\calM$ & pricing operator & $\mathbf y$ & output weights \\
$\Jop u=\E[u(x+J)]$ & expected value after a jump & $N_c,N_0,N_b$ & numbers of interior, initial, boundary points \\
$\lambda,\ g,\ \kappa_J$ & jump rate, log-jump density, $\E[e^J-1]$ & $w_{\mathrm{ic}},\ w_{\mathrm b}$ & weights of initial and boundary conditions \\
$v,\ \vbar,\ \kappa,\ \xi$ & Heston variance, long-run level, speed, vol of var & $\reg$ & ridge (Tikhonov) parameter \\
$F,\ \alpha,\ \beta,\ \nu$ & SABR forward, volatility, exponent, vol of vol & $\theta,\ \npar$ & model-parameter vector and its length \\
$\rho$ & correlation of the two Brownian motions & $\nens$ & number of independent feature draws averaged \\
$\psi$ & boundary data & $u_{\mathrm{BS}}$ & Black--Scholes price \\
$r,\ q$ & interest rate, dividend yield & & \\
\bottomrule
\end{tabular}
\end{table}

\section{The method}\label{sec:method}

\subsection{Random probit features}\label{sec:ansatz}

Let $z$ denote the vector of state variables and time: $z=(x,\tau)$ for the
jump models, $z=(x,v,\tau)$ for Heston and $z=(F,\alpha,\tau)$ for SABR,
and let $d$ be its length. We approximate the price by a linear
combination of $\nfeat$ functions of the form $\Phi(A_k\cdot(z-\zeta_k))$.
For $k=1,\dots,\nfeat$ we draw, independently,
\begin{itemize}[leftmargin=*]
\item an inner weight vector $A_k\in\R^d$ whose components are independent
  and normal, $A_k^{(j)}\sim\calN(0,\omega_j^2)$ for $j=1,\dots,d$, where
  the bandwidth $\omega_j>0$ is a fixed choice for each coordinate; and
\item a centre $\zeta_k$, uniformly distributed over the computational
  domain.
\end{itemize}
These draws are made once and then held fixed. The \emph{pre-activation}
of feature $k$ is
\begin{equation*}
s_k(z) := A_k\cdot(z-\zeta_k),
\end{equation*}
and the feature is $\Phi(s_k(z))$. The pre-activation is zero at the
centre $\zeta_k$, and more generally at every point $z$ of the hyperplane
through $\zeta_k$ perpendicular to $A_k$. On this hyperplane the feature
equals $\Phi(0)=\tfrac12$. Moving away from it in the direction of $A_k$,
the feature rises towards $1$; moving in the opposite direction, it falls
towards $0$. The centre therefore fixes where the transition takes place,
and $A_k$ fixes its direction and steepness: in coordinate $j$ the width of
the transition is about $1/\abs{A_k^{(j)}}$. Large bandwidths $\omega_j$
give sharp features and small bandwidths give smooth ones.

\begin{definition}[Random-feature approximation]\label{def:ansatz}
For output weights $\mathbf{y}=(y_1,\dots,y_\nfeat)\in\R^{\nfeat}$,
\begin{equation}\label{eq:ansatz}
\widehat V(z;\mathbf{y}) := \sum_{k=1}^{\nfeat} y_k\,\Phi\big(s_k(z)\big).
\end{equation}
\end{definition}

Only the output weights $\mathbf y$ are determined from the pricing
problem. In neural-network terms, \eqref{eq:ansatz} is a network with one
hidden layer of $\nfeat$ neurons, whose inner weights are sampled and
never trained.

We use the probit activation $\Phi$ for three reasons. First, the
approximation \eqref{eq:ansatz} is flexible enough. Random neural
networks, in which the inner weights are drawn at random and only the
output weights are fitted, have the universal approximation property
under mild conditions on the activation \chgg{(it must not be a polynomial
and must be bounded with bounded derivatives, or a limit of such
functions in a weighted norm), which the probit function satisfies}
\citep[Corollary~3.8]{neufeldschmocker2023}, and
for suitable target functions the $L^2$ approximation error decreases
like $\nfeat^{-1/2}$ \citep{rahimi2008}.
Second,
$\Phi$ is infinitely differentiable, and its derivatives are explicit:
$\Phi'=\phi$ and $\phi'(s)=-s\,\phi(s)$. Third, and most important for the
jump models, averaging a probit feature over a normal or a
double-exponential jump gives an explicit formula
(Lemma~\ref{lem:jump}). Because $\Phi$ is smooth, however, a finite sum of
probit features cannot reproduce the kink of the call payoff exactly; the
consequences are discussed in Sections~\ref{sec:bs-subtraction}
and~\ref{sec:convergence}.

\subsection{Fitting by least squares}\label{sec:lsq}

\label{sec:boundary}The pricing
equations of Section~\ref{sec:models} are posed on unbounded domains. We
solve them on a bounded box, for example $x\in[x_{\min},x_{\max}]$ for the
jump models, and impose boundary data $\psi$ on some of its faces. The
faces and the data used for each model are given with the experiments in
Section~\ref{sec:experiments}. \chg{In all calibrations the data $\psi$ do
not depend on the model parameters, which lets the calibration Jacobian of
Section~\ref{sec:calibration} take its simplest form. In several
experiments, conditions on some faces are omitted altogether; the
least-squares fit is then determined there by the equation and the
payoff.}

We choose $\mathbf y$ so that $\widehat V$ satisfies, as closely as
possible, the three conditions that define the price: the pricing
equation $\calM[\Vb]=0$ inside the domain, the payoff $\Vb(\cdot,0)=f$ at
$\tau=0$, and the boundary data $\psi$ on the boundary faces. Each condition is imposed at a finite set of points, the
\emph{collocation points}:
\begin{itemize}[leftmargin=*]
\item $N_c$ interior points $z_1,\dots,z_{N_c}$ in the domain with
  $\tau>0$;
\item $N_0$ initial points $x^{(0)}_1,\dots,x^{(0)}_{N_0}$ at $\tau=0$, a
  fraction of them clustered near the strike $x=0$, where the payoff has
  its kink;
\item $N_b$ boundary points $z^{(b)}_1,\dots,z^{(b)}_{N_b}$ on the faces
  where boundary data are imposed.
\end{itemize}
Unless an experiment states otherwise, the interior and boundary points
are drawn uniformly at random. The loss is the weighted sum of the squared
errors in the three conditions:
\begin{equation}\label{eq:loss}
L(\mathbf{y}) :=
  \sum_{i=1}^{N_c}\big|\calM[\widehat V](z_i;\mathbf{y})\big|^2
  + w_{\mathrm{ic}}^2\sum_{j=1}^{N_0}\big|\widehat V(x_j^{(0)},0;\mathbf{y})-f(x_j^{(0)})\big|^2
  + w_{\mathrm{b}}^2\sum_{\ell=1}^{N_b}\big|\widehat V(z_\ell^{(b)};\mathbf{y})-\psi_\ell\big|^2 ,
\end{equation}
where $\psi_\ell$ is the boundary value at the point $z^{(b)}_\ell$. All
three terms are sums over the collocation points, so the balance between
them depends both on the weights and on $N_c$, $N_0$ and $N_b$, which we
report with each experiment. The weights appear squared because, in the
matrix form below, each initial or boundary row is multiplied by
$w_{\mathrm{ic}}$ or $w_{\mathrm b}$.

\label{sec:weights}\chgb{The choice of the weights decides}
how the solve trades the three conditions against each other, since a
fixed number of output weights must serve all of them. With a small
$w_{\mathrm{ic}}$ the solution fits the equation closely but the payoff
less well, and the error at $\tau=0$ then spreads to the whole price
surface; with a large $w_{\mathrm{ic}}$ the opposite happens. A small
error in the equation alone therefore does not guarantee a small price
error. \chg{In our experiments the accuracy depended on the weights only
mildly over a moderate range, and deteriorated only when one condition
was weighted far more or far less than the others.}

Since $\calM$ is linear, $\calM[\widehat V](z;\mathbf y)=\sum_k
y_k\,\calM[\Phi(s_k)](z)$, and every term in \eqref{eq:loss} is a linear
function of $\mathbf y$ minus a target. The loss is therefore a linear
least-squares objective,
\[
L(\mathbf y)=\norm{\mathbf A\mathbf y-\mathbf b}_2^2,
\qquad \mathbf A\in\R^{N\times\nfeat},\quad N=N_c+N_0+N_b,
\]
in which each collocation point contributes one row of $\mathbf A$ and one
entry of $\mathbf b$:
\begin{itemize}[leftmargin=*]
\item an interior point $z_i$ gives the row
  $\big(\calM[\Phi(s_k)](z_i)\big)_{k=1}^{\nfeat}$ with target $0$;
\item an initial point gives the row
  $w_{\mathrm{ic}}\,\big(\Phi(s_k(x_j^{(0)},0))\big)_{k}$ with target
  $w_{\mathrm{ic}}\,f(x_j^{(0)})$;
\item a boundary point gives the row
  $w_{\mathrm b}\,\big(\Phi(s_k(z_\ell^{(b)}))\big)_k$ with target
  $w_{\mathrm b}\,\psi_\ell$.
\end{itemize}
The only entries that need work are $\calM[\Phi(s_k)](z_i)$, the pricing
operator applied to a single feature; Sections~\ref{sec:general-op}
and~\ref{sec:merton-formulas} show that they are explicit.

The matrix $\mathbf A$ is typically very ill-conditioned, because many
random features are nearly linear combinations of others. We therefore
add a small ridge (Tikhonov) penalty, which selects, among the many weight
vectors that fit almost equally well, one of small norm:
\begin{equation}\label{eq:lsq-reg}
\mathbf{y}^\ast
 = \argmin_{\mathbf{y}\in\R^{\nfeat}}\Big\{\norm{\mathbf A\mathbf{y}-\mathbf b}_2^2+\reg\norm{\mathbf{y}}_2^2\Big\}
 = \mathbf H^{-1}\mathbf A^\top\mathbf b,
\qquad \mathbf H := \mathbf A^\top\mathbf A+\reg I .
\end{equation}
The matrix $\mathbf H$ is symmetric and positive definite, and we solve
\eqref{eq:lsq-reg} with its Cholesky factorisation. \chgb{This writes
$\mathbf H=\mathbf L\mathbf L^\top$, where $\mathbf L$ is lower
triangular, that is, all its entries above the diagonal are zero, and it
costs about $\nfeat^3/3$ operations. Once $\mathbf L$ is known, a system
$\mathbf H\mathbf z=\mathbf r$ with any other right-hand side $\mathbf r$
is solved in two cheap passes: first $\mathbf L\mathbf w=\mathbf r$ from
the top down, each row giving one more unknown, and then $\mathbf
L^\top\mathbf z=\mathbf w$ from the bottom up. These two triangular solves
cost about $2\nfeat^2$ operations together; for $\nfeat=1800$ that is
about $300$ times less than the factorisation.} This factorisation is
reused for all sensitivities in Section~\ref{sec:calibration}\chgb{, and
for any other payoff under the same model}. The
fitted price is $\widehat V(z;\mathbf y^\ast)$, and the price in currency
units is $K\,\widehat V$.

\label{sec:algorithm}\chgb{The cost of one solve has three parts:} assembling $\mathbf A$ costs
$O(N\nfeat)$ evaluations of explicit functions, forming $\mathbf
A^\top\mathbf A$ costs $O(N\nfeat^2)$ operations and the Cholesky
factorisation $O(\nfeat^3)$. For the sizes used in the experiments
\chgb{($\nfeat$ from \chgg{$800$ to $1800$} for the jump models to several thousand
for Heston and SABR) one solve takes seconds to minutes on a single
machine}. Different draws of the features give slightly
different prices, so we report results averaged over $\nens$ independent
draws, with $\nens$ stated for each experiment. \chgb{The spread of the
prices over the draws also indicates the size of the error when no
reference price is available (Section~\ref{sec:heston-exp})}\chgo{, but
only of the part of the error that varies between draws
(Section~\ref{sec:identifiability})}\chgb{.}

\chgb{The factorisation can also be reused for other payoffs. The
matrix $\mathbf A$ depends on the model, its parameters and the sample
points, but not on the payoff: the payoff enters only the targets of the
initial and boundary rows. Another European payoff under the same model
is therefore priced with the same factorisation, at the cost of one
product with $\mathbf A^\top$ and two triangular solves, and gives its
whole price surface and Greeks. How accurate that price is depends on how
well the features resolve the new payoff's kinks or jumps; the features
and sample points used here are concentrated at the kink of a call, and
we test only calls. With the Black--Scholes subtraction of
Section~\ref{sec:bs-subtraction}, the subtracted function must also match
the new payoff. The same reuse is possible for any linear
discretisation, including finite differences; the difference is in the
cost. A time-stepping scheme must repeat its whole march in time for each
new payoff, whereas here the expensive steps, assembling and factorising
the matrix, are done once.}

\subsection{The local part of the operator}\label{sec:general-op}

We first apply the local part of the pricing operator \eqref{eq:M-split}
to a single feature; the jump term of the Merton and Kou models is treated
in Section~\ref{sec:merton-formulas}. Only two facts are needed: the chain
rule for $\Phi(s_k)$, with $\phi'(s)=-s\,\phi(s)$, and the values of the
coefficients at the point. Writing $A_k^{(j)}$ for the component of $A_k$
along the coordinate $z_j$,
\begin{equation}\label{eq:feature-derivs}
\partial_{z_j}\Phi(s_k)=A_k^{(j)}\phi(s_k),\qquad
\partial_{z_i}\partial_{z_j}\Phi(s_k)=-A_k^{(i)}A_k^{(j)}\,s_k\,\phi(s_k).
\end{equation}
In the statement below, $\iota=(\iota_1,\dots,\iota_d)$ is a multi-index,
$\partial^\iota=\partial_{z_1}^{\iota_1}\cdots\partial_{z_d}^{\iota_d}$, and
$A_k^\iota=\prod_j (A_k^{(j)})^{\iota_j}$.

\begin{lemma}[Local operator applied to a feature]\label{lem:pointwise}
Let
\[
\calM_{\mathrm{loc}}[u] = \sum_{1\le|\iota|\le 2} a_\iota(z)\,\partial^{\iota}u + c(z)\,u
\]
be a linear operator of order at most two whose coefficients are finite at
the point $z$. Then
\begin{equation}\label{eq:pointwise}
\calM_{\mathrm{loc}}[\Phi(s_k)](z) = c(z)\,\Phi(s_k) + \big(P_k(z)\,s_k+Q_k(z)\big)\,\phi(s_k),
\end{equation}
with
\begin{equation}\label{eq:PkQk}
P_k(z) = -\sum_{|\iota|=2} a_\iota(z)\,A_k^{\iota},
\qquad
Q_k(z) = \sum_{|\iota|=1} a_\iota(z)\,A_k^{\iota}.
\end{equation}
\end{lemma}

\begin{proof}
By the chain rule, $\partial^{\iota}\Phi(s_k)=A_k^{\iota}\,\phi(s_k)$ for
$|\iota|=1$ and $\partial^{\iota}\Phi(s_k)=-A_k^{\iota}\,s_k\,\phi(s_k)$
for $|\iota|=2$. Summing over $\iota$ gives \eqref{eq:pointwise}.
\end{proof}
The operator applied to a feature is the reaction coefficient
times the feature, plus the density $\phi(s_k)$ times a linear function of
$s_k$. The coefficient $P_k$ collects the second-order terms, and $Q_k$
the first-order terms, including the time derivative. \chg{The result is an
application of the chain rule and holds for any smooth activation (tanh,
for example) with the appropriate derivatives in place of $\phi$ and
$-s\phi$; what is particular to the probit is the jump term of
Lemma~\ref{lem:jump}.} Because the coefficients enter only through their
values at the point, state-dependent coefficients cost no more than
constant ones.

We now give the coefficients for each model, read off from
\eqref{eq:M-jump}, \eqref{eq:M-heston} and \eqref{eq:M-sabr}. Here
$A_k^{(x)}$, $A_k^{(v)}$, $A_k^{(F)}$, $A_k^{(\alpha)}$ and $A_k^{(\tau)}$
denote the components of $A_k$ along $x$, $v$, $F$, $\alpha$ and $\tau$.

\chgb{For Merton and Kou, with} $z=(x,\tau)$, the reaction coefficient
is $c=r+\lambda$, and the two coefficients are constants:
\begin{equation}\label{eq:PkQk-merton}
P_k = \tfrac12\sigma^2\big(A_k^{(x)}\big)^2,
\qquad
Q_k = A_k^{(\tau)}-\mu_0\,A_k^{(x)} .
\end{equation}

\chgb{For Heston, with} $z=(x,v,\tau)$, the reaction coefficient is
$c=r$ and
\begin{equation}
	\left\{\begin{array}{l}
P_k(v) = \tfrac12 v\big(A_k^{(x)}\big)^2
        + \rho\xi v\,A_k^{(x)}A_k^{(v)}
        + \tfrac12\xi^2 v\,\big(A_k^{(v)}\big)^2,
        \\ [6pt]
Q_k(v) = A_k^{(\tau)}
        - \big(r-q-\tfrac{v}{2}\big)A_k^{(x)}
        - \kappa(\vbar-v)\,A_k^{(v)}.
\end{array}\right. \label{eq:PkQk-heston} 
\end{equation}
Both are affine in $v$.

\chgb{For SABR, with} $z=(F,\alpha,\tau)$, there is no discounting, so
$c=0$, and
\begin{equation}
	\left\{\begin{array}{l}
P_k(F,\alpha) = \tfrac12\alpha^2 F^{2\beta}\big(A_k^{(F)}\big)^2
        + \rho\nu\alpha^{2} F^{\beta}A_k^{(F)}A_k^{(\alpha)}
     + \tfrac12\nu^2\alpha^2\big(A_k^{(\alpha)}\big)^2,\\ [6pt]
Q_k = A_k^{(\tau)}.
\end{array}\right. \label{eq:PkQk-sabr} 
\end{equation}
The nonlinear dependence on $F^\beta$ enters only through the value of
$P_k$ at the collocation point.

\subsection{The jump term}\label{sec:merton-formulas}

For the Merton and Kou models the pricing operator \eqref{eq:M-jump} also
contains the jump term $-\lambda\,\Jop u$.
A jump changes $x$ to $x+J$, and hence changes
the pre-activation from $s_k$ to $s_k+A_k^{(x)}J$. The jump term of
\eqref{eq:M-jump} applied to a feature is therefore
\[
\Jop\Phi(s_k) = \E\big[\Phi\big(s_k + A_k^{(x)}J\big)\big],
\]
an expectation over the jump size only. The next lemma gives it in closed
form for both jump laws. Part~(i) is a classical identity for normal
integrals \citep{owen1980}; we include its short proof for completeness.

\begin{lemma}[Explicit jump term]\label{lem:jump}
Let $s\in\R$ and $a\in\R$.
\begin{enumerate}[label=(\roman*),leftmargin=*]
\item If $J\sim\calN(\mu_J,\sigma_J^2)$ (Merton), then
\begin{equation}\label{eq:probit-merton}
\E\big[\Phi(s+aJ)\big]
 = \Phi\!\left(\frac{s + a\mu_J}{\sqrt{1+a^2\sigma_J^2}}\right).
\end{equation}
\item If $J$ has the double-exponential density \eqref{eq:g-kou} (Kou), then
\begin{equation}\label{eq:probit-kou-sum}
\E\big[\Phi(s+aJ)\big] = p\,G(s,a,\eta_1)+(1-p)\,G(s,-a,\eta_2),
\end{equation}
where, for $a\ne0$ and $\eta>0$,
\begin{equation}\label{eq:probit-kou}
G(s,a,\eta) := \E\big[\Phi(s+aY)\big]
 = \Phi(s) + \tfrac12\,\mathrm{sgn}(a)\,e^{-s^2/2}\,
              \operatorname{erfcx}\!\Big(\frac{a s+\eta}{\sqrt2\,|a|}\Big),
\qquad Y\sim\mathrm{Exp}(\eta),
\end{equation}
$\operatorname{erfcx}(u)=e^{u^2}\operatorname{erfc}(u)$ is the scaled
complementary error function, and $G(s,0,\eta)=\Phi(s)$.
\end{enumerate}
\end{lemma}

\begin{proof}
(i) Let $W$ be a standard normal variable independent of $J$, so that
$\Phi(u)=\Prob(W\le u)$. Then
$\E[\Phi(s+aJ)]=\Prob(W\le s+aJ)=\Prob(W-aJ\le s)$. The variable $W-aJ$ is
normal with mean $-a\mu_J$ and variance $1+a^2\sigma_J^2$, which gives
\eqref{eq:probit-merton}.

(ii) \chg{The Kou jump is $J=Y_1$ with probability $p$ and $J=-Y_2$ with
probability $1-p$, where $Y_i\sim\mathrm{Exp}(\eta_i)$. Conditioning on the
direction of the jump gives \eqref{eq:probit-kou-sum}, since
$\E[\Phi(s-aY_2)]=G(s,-a,\eta_2)$. It remains to compute $G$. Integrating
by parts,
\[
G(s,a,\eta)=\int_0^\infty \eta e^{-\eta y}\,\Phi(s+ay)\,dy
 = \Phi(s)+a\int_0^\infty e^{-\eta y}\,\phi(s+ay)\,dy .
\]
To compute the last integral, write the integrand as one exponential:
\[
e^{-\eta y}\,\phi(s+ay)=\frac{1}{\sqrt{2\pi}}
\exp\!\Big(-\big[\eta y+\tfrac12(s+ay)^2\big]\Big).
\]
The bracket is a quadratic in $y$, namely
$\tfrac12a^2y^2+(as+\eta)y+\tfrac12 s^2$. Completing the square, with
\[
m=\frac{as+\eta}{a^2},\qquad q=\frac{as+\eta}{|a|},\qquad
\text{so that}\quad \tfrac12a^2m^2=\tfrac12q^2,
\]
gives $\eta y+\tfrac12(s+ay)^2=\tfrac12a^2(y+m)^2+\tfrac12s^2-\tfrac12q^2$,
and hence
\[
e^{-\eta y}\,\phi(s+ay)=e^{(q^2-s^2)/2}\,\frac{1}{\sqrt{2\pi}}\,
e^{-\frac12a^2(y+m)^2}.
\]
The first factor does not depend on $y$. In the remaining integral we
substitute $w=|a|(y+m)$, so that $a^2(y+m)^2=w^2$ and $dy=dw/|a|$; the
lower limit $y=0$ becomes $w=|a|m=q$, and the upper limit stays at
$+\infty$ for either sign of $a$. Therefore
\[
a\int_0^\infty e^{-\eta y}\,\phi(s+ay)\,dy
 = \frac{a}{|a|}\,e^{(q^2-s^2)/2}\int_q^\infty \phi(w)\,dw
 = \mathrm{sgn}(a)\,e^{(q^2-s^2)/2}\,\big(1-\Phi(q)\big).
\]
Finally, $1-\Phi(q)=\tfrac12 e^{-q^2/2}\operatorname{erfcx}(q/\sqrt2)$;
the factors $e^{\pm q^2/2}$ cancel, which gives \eqref{eq:probit-kou}.
Because the substitution uses $|a|$, the calculation covers both signs of
$a$.} Both identities were also checked against numerical quadrature, to
machine precision.
\end{proof}

With $a=A_k^{(x)}$ and $s=s_k$, Lemma~\ref{lem:jump} gives $\Jop\Phi(s_k)$
in closed form. Adding it to the local part (Lemma~\ref{lem:pointwise}
with the coefficients \eqref{eq:PkQk-merton}), the Merton operator
applied to feature $k$ is
\begin{equation}\label{eq:M-merton-feature}
\calM[\Phi(s_k)]
 = (r+\lambda)\,\Phi(s_k) + \big(P_k\,s_k+Q_k\big)\,\phi(s_k)
   - \lambda\,\Phi\!\left(\frac{s_k + A_k^{(x)}\mu_J}{\sqrt{1+(A_k^{(x)})^2\sigma_J^2}}\right).
\end{equation}
For Kou the last term is replaced by $\lambda$ times
\eqref{eq:probit-kou-sum}. Every entry of the interior rows of $\mathbf A$
is therefore an explicit expression in $\Phi$, $\phi$ and
$\operatorname{erfcx}$, evaluated at the collocation point, and no numerical integration over the
jump size is needed.

\begin{remark}[Fourier features]\label{sec:fourier}
The probit is not the only activation for which both the local part and
the jump term are explicit. With the pair of features $\cos s_k$ and
$\sin s_k$ in place of $\Phi(s_k)$, every derivative of a feature is
again a multiple of $\cos s_k$ or $\sin s_k$, and the jump term follows
from the characteristic function $\varphi_J(u)=\E[e^{iuJ}]$ of the jump:
\[
\E\big[\cos(s_k+aJ)\big]+i\,\E\big[\sin(s_k+aJ)\big] = e^{is_k}\,\varphi_J(a).
\]
For Merton, for example,
$\E[\cos(s_k+aJ)]=e^{-a^2\sigma_J^2/2}\cos(s_k+a\mu_J)$. The residual is
again linear in the output weights, and the method applies unchanged.
These features oscillate over the whole domain and suit smooth solutions;
Section~\ref{sec:bansal} uses them on smooth manufactured test problems,
where they are more accurate than probit features. A probit feature is a
single smooth step, which better matches the local shape of a payoff
kink, and we use probit features for the option-pricing problems.
\end{remark}

\subsection{Subtracting the Black--Scholes price}\label{sec:bs-subtraction}

The call payoff $f(x)=(e^x-1)^+$ has a kink at the strike $x=0$. At short
maturities the price is still close to this kink, and a finite sum of
smooth features cannot reproduce it well. The error of the plain solve is
then concentrated near the strike at small $\tau$, and it stops
decreasing as features are added (Section~\ref{sec:convergence}). For the
jump models we remove the difficulty by subtracting a known function that
already contains the kink, and solving for the remainder. This is the
classical technique of singularity subtraction \citep{fix1973};
in option pricing the kink is more often handled by smoothing the payoff
or by special start-up steps of a time-stepping scheme \citep{pooley2003}.

\chgb{In general, let} $h$ be any function for which
$\calM[h]$ can be evaluated explicitly; for the jump models this includes
$\Jop h$. Writing $\Vb=h+w$ and using $\calM[\Vb]=0$ and the linearity of
$\calM$, the remainder $w$ solves
\begin{equation}\label{eq:lifting}
\calM[w] = -\calM[h],
\qquad
w(\cdot,0) = f - h(\cdot,0),
\qquad
w = \psi - h \ \text{on the boundary.}
\end{equation}
This is a problem of the same type as before, with an explicit right-hand
side, so it is solved by the same least-squares method; the only change is
that the interior targets in $\mathbf b$ become $-\calM[h](z_i)$
instead of $0$. The decomposition is exact and assumes nothing about $w$.
It improves the accuracy when $w$ is easier to approximate than $\Vb$.

We take
$h=u_{\mathrm{BS}}$, and, as in all our jump-model experiments, set $q=0$
in this subsection. Here $u_{\mathrm{BS}}$ is the normalised Black--Scholes
call price with the
same strike, interest rate and diffusion volatility $\sigma$:
\[
u_{\mathrm{BS}}(x,\tau) = e^x\,\Phi(d_1) - e^{-r\tau}\Phi(d_2),
\qquad
d_{1,2}=\frac{x+(r\pm\frac12\sigma^2)\tau}{\sigma\sqrt\tau}.
\]
It has the same payoff, so the remainder starts from zero,
$w(x,0)=f(x)-u_{\mathrm{BS}}(x,0)=0$. It also satisfies the
Black--Scholes equation
$\partial_\tau u - \tfrac12\sigma^2\partial_{xx}u-(r-\tfrac12\sigma^2)\partial_xu+ru=0$,
which is the Merton operator \eqref{eq:M-jump} without the jump terms.
Substituting into \eqref{eq:lifting}, only the jump terms remain on the
right-hand side:
\begin{equation}\label{eq:bs-source}
\calM[w] = \calF(x,\tau) := \lambda\Big(\Jop u_{\mathrm{BS}}(x,\tau)
  - u_{\mathrm{BS}}(x,\tau) - \kappa_J\,\partial_x u_{\mathrm{BS}}(x,\tau)\Big),
\end{equation}
where $\partial_x u_{\mathrm{BS}}=e^x\Phi(d_1)$ is the normalised
Black--Scholes delta.

The jump average $\Jop u_{\mathrm{BS}}$ is explicit for both jump laws.
For Merton it is again a Black--Scholes-type price: since $x+J$ plus the
Black--Scholes log-return is normal with mean
$x+m$, $m=(r-\tfrac12\sigma^2)\tau+\mu_J$, and variance
$q^2=\sigma^2\tau+\sigma_J^2$,
\begin{equation}\label{eq:JuBS-merton}
\Jop u_{\mathrm{BS}}(x,\tau)
 = e^{-r\tau}\Big(e^{x+m+q^2/2}\,\Phi\Big(\frac{x+m+q^2}{q}\Big)
   - \Phi\Big(\frac{x+m}{q}\Big)\Big).
\end{equation}
\chg{For Kou it is expressed through the function $G$ of
Lemma~\ref{lem:jump}. With $b=1/(\sigma\sqrt\tau)$,
\begin{equation}\label{eq:JuBS-kou}
\Jop u_{\mathrm{BS}}(x,\tau)
 = e^x\Big(\tfrac{p\eta_1}{\eta_1-1}\,G(d_1,b,\eta_1-1)
         + \tfrac{(1-p)\eta_2}{\eta_2+1}\,G(d_1,-b,\eta_2+1)\Big)
 - e^{-r\tau}\Big(p\,G(d_2,b,\eta_1)+(1-p)\,G(d_2,-b,\eta_2)\Big),
\end{equation}
which follows from the Kou case of Lemma~\ref{lem:jump} after absorbing
the factor $e^{\pm Y}$ into the exponential density; this needs
$\eta_1>1$, which is already required for $\kappa_J$ to be finite.} We
checked \eqref{eq:JuBS-merton} and \eqref{eq:JuBS-kou} against numerical
quadrature. The jump identity is thus used twice: once for the jump term
$\Jop w$ of the unknown remainder, through the features, and once for the
right-hand side $\calF$. No quadrature enters either.

\chg{The remainder is smoother than the price, but it is not free of
structure near the strike. As $\tau\to0$ the right-hand side
\eqref{eq:bs-source} tends to $\lambda(\Jop f - f - \kappa_J f')$, and
$f'(x)=e^x\mathbf 1_{\{x>0\}}$ jumps at the strike. The remainder therefore
has a thin layer near $x=0$ at short maturities, of width about
$\sigma\sqrt\tau$.} To resolve it, we devote a fraction of the features to
this region: their centres are concentrated near $x=0$ and at small
$\tau$, and their slopes $A_k^{(x)}$ are of order $1/(\sigma\sqrt\tau)$
over a geometric sequence of time scales. These are ordinary probit
features, with the same explicit jump term, so the solve is unchanged.
The ablation in Section~\ref{sec:merton-exp} measures the effect of each
step: \chgb{for Merton,} the subtraction alone reduces the error by a
factor of about \chgg{$100$}, the extra features alone by about \chgg{$7$}, and the two
together by about \chgg{$700$} \chgb{(for Kou, \chgg{$70$, $15$ and $300$})}, \chg{with the total number of features held fixed in all four
runs}. \chgg{Nothing in the subtraction is specific to random features:
applied to the finite-difference scheme of Section~\ref{sec:merton-exp},
it reduces that scheme's error by a factor of about $200$ for Merton and
$70$ for Kou.}

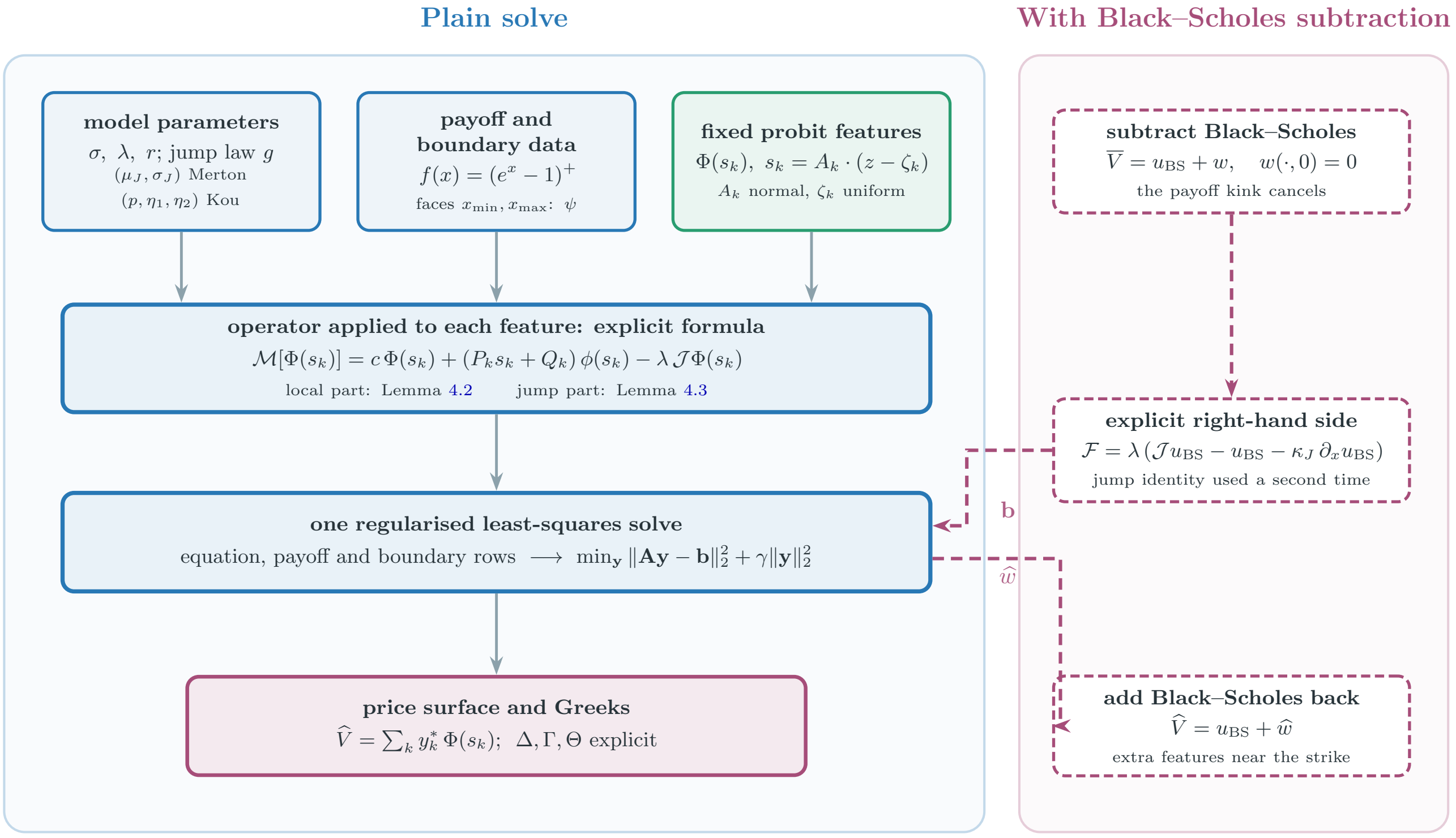
\begin{figure*}[!htbp]
\centering
\resizebox{\textwidth}{!}{%
\begin{tikzpicture}[
  font=\small,
  >={Stealth[length=2.8mm,width=2mm]},
  box/.style={rounded corners=5pt, draw=mut!60, fill=white, line width=1pt,
              align=center, inner sep=6pt, minimum height=15mm, text=ink},
  inbox/.style={box, draw=oiblue, fill=oiblue!8, line width=1.3pt, text width=38mm, minimum height=21mm},
  featbox/.style={box, draw=oigreen, fill=oigreen!10, line width=1.3pt, text width=38mm, minimum height=21mm},
  shared/.style={box, draw=oiblue, fill=oiblue!10, line width=1.8pt, text width=128mm},
  result/.style={box, draw=oiverm, fill=oiverm!12, line width=1.6pt, text width=90mm},
  carrier/.style={box, draw=oiverm, fill=white, line width=1.3pt, dashed,
                  dash pattern=on 4pt off 2pt, text width=50mm},
  flow/.style={->, line width=1.2pt, draw=mut!85},
  v2/.style={->, line width=1.5pt, draw=oiverm, dashed, dash pattern=on 5pt off 3pt},
  tag/.style={font=\large\bfseries},
  lbl/.style={font=\footnotesize\itshape, text=oiverm!90, align=center},
]
\path (0,3.0) coordinate (tL) (0,-8.6) coordinate (bL);
\path (21.0,3.0) coordinate (tR) (21.0,-8.6) coordinate (bR);
\node[inbox]   (par) at (2.2, 1.5) {\textbf{model parameters}\\[2pt]
   \mbox{$\sigma,\ \lambda,\ r$; jump law $g$}\\[1pt]
   \scriptsize \mbox{$(\mu_J,\sigma_J)$ Merton}\\ \scriptsize \mbox{$(p,\eta_1,\eta_2)$ Kou}};
\node[inbox]   (pay) at (7.0, 1.5) {\textbf{payoff and boundary data}\\[2pt]
   \mbox{$f(x)=(e^x-1)^+$}\\[1pt] \scriptsize \mbox{faces $x_{\min},x_{\max}$: $\psi$}};
\node[featbox] (fb)  at (11.8,1.5) {\textbf{fixed probit features}\\[2pt]
   \mbox{$\Phi(s_k),\ s_k=A_k\cdot(z-\zeta_k)$}\\[1pt] \scriptsize \mbox{$A_k$ normal, $\zeta_k$ uniform}};
\node[shared] (op) at (7.0,-1.5) {\textbf{operator applied to each feature: explicit formula}\\[3pt]
   \mbox{$\calM[\Phi(s_k)]=c\,\Phi(s_k)+(P_k s_k+Q_k)\,\phi(s_k)-\lambda\,\Jop\Phi(s_k)$}\\[2pt]
   \scriptsize local part: Lemma~\ref{lem:pointwise} \qquad jump part: Lemma~\ref{lem:jump}};
\node[shared] (ls) at (7.0,-4.3) {\textbf{one regularised least-squares solve}\\[3pt]
   \mbox{equation, payoff and boundary rows $\;\longrightarrow\;$
   $\min_{\mathbf y}\|\mathbf A\mathbf y-\mathbf b\|_2^2+\reg\|\mathbf y\|_2^2$}};
\node[result] (out) at (7.0,-7.1) {\textbf{price surface and Greeks}\\[3pt]
   \mbox{$\widehat V=\textstyle\sum_k y_k^\ast\,\Phi(s_k)$; \ $\Delta,\Gamma,\Theta$ explicit}};
\draw[flow] (par.south) -- (par.south |- op.north);
\draw[flow] (pay.south) -- (pay.south |- op.north);
\draw[flow] (fb.south)  -- (fb.south  |- op.north);
\draw[flow] (op) -- (ls);
\draw[flow] (ls) -- (out);
\node[carrier] (sub) at (18.2, 1.5) {\textbf{subtract Black--Scholes}\\[2pt]
   \mbox{$\Vb=u_{\mathrm{BS}}+w,\quad w(\cdot,0)=0$}\\[1pt]
   \scriptsize the payoff kink cancels};
\node[carrier] (src) at (18.2,-2.9) {\textbf{explicit right-hand side}\\[2pt]
   \mbox{$\calF=\lambda\,(\Jop u_{\mathrm{BS}}-u_{\mathrm{BS}}-\kappa_J\,\partial_x u_{\mathrm{BS}})$}\\[1pt]
   \scriptsize jump identity used a second time};
\node[carrier] (add) at (18.2,-7.1) {\textbf{add Black--Scholes back}\\[2pt]
   \mbox{$\widehat V=u_{\mathrm{BS}}+\widehat w$}\\[1pt]
   \scriptsize \mbox{extra features near the strike}};
\draw[v2] (sub) -- (src);
\draw[v2] (src.west) -- ++(-1.3,0) |- ($(ls.east)+(0,0.25)$);
\node[lbl, anchor=south, font=\normalsize] at ($(ls.east)+(1.15,0.25)$) {$\mathbf b$};
\draw[v2] ($(ls.east)+(0,-0.25)$) -- ($(ls.east)+(1.95,-0.25)$) |- (add.west);
\node[lbl, anchor=north, font=\normalsize] at ($(ls.east)+(1.15,-0.25)$) {$\widehat w$};
\begin{scope}[on background layer]
  \node[rounded corners=9pt, fill=oiblue!3, draw=oiblue!28, line width=1pt,
        fit=(par)(fb)(out)(op)(ls)(tL)(bL), inner xsep=5mm] (Pv1) {};
  \node[rounded corners=9pt, fill=oiverm!3, draw=oiverm!28, line width=1pt,
        fit=(sub)(add)(tR)(bR), inner xsep=5mm] (Pv2) {};
\end{scope}
\node[tag, text=oiblue] at (Pv1.north) [above=3mm] {Plain solve};
\node[tag, text=oiverm] at (Pv2.north) [above=3mm] {With Black--Scholes subtraction};
\end{tikzpicture}}%
\caption{The two pipelines for the jump models. Left: the plain solve.
Right: the solve with Black--Scholes subtraction, which solves the same
kind of system for the remainder $w=\Vb-u_{\mathrm{BS}}$, with the
explicit right-hand side $\calF$ of \eqref{eq:bs-source}.}
\label{fig:mert-flow}
\end{figure*}

Figure~\ref{fig:mert-flow} shows the two pipelines side by side. In the
experiments we refer to them as the \emph{plain solve} and the
\emph{solve with Black--Scholes subtraction}.

\chgo{The same step can be used for Heston and SABR, with a different
choice of $h$: the price of the same model with the volatility of
volatility set to zero. In the jump models, $u_{\mathrm{BS}}$ is the price
with jump intensity zero, so this is the same rule. With $\xi=0$, the
Heston variance follows the deterministic path
$dv_t=\kappa(\vbar-v_t)\,dt$, and $h$ is the Black--Scholes price in which
$\sigma^2\tau$ is replaced by the integrated variance
\[
W(v,\tau)=\vbar\,\tau+(v-\vbar)\,\frac{1-e^{-\kappa\tau}}{\kappa}.
\]
With $\nu=0$ and $\beta=1$, the SABR volatility stays at $\alpha$, and $h$
is Black's price with $W=\alpha^2\tau$. These are the leading terms of
the small volatility-of-volatility expansions of \citet{lewis2000} and
\citet{benhamou2010}; here they serve only as the known function in
\eqref{eq:lifting}. As before, $h$ has the payoff of the option, so $w$
starts from zero, and $h$ solves the pricing equation with the volatility
of volatility set to zero, so only the terms that contain it remain on
the right-hand side. For Heston,
\[
-\calM[h]=\rho\xi v\,W_v\,\partial_{xW}h+\tfrac12\xi^2 v\,W_v^2\,\partial_{WW}h,
\qquad W_v=\frac{1-e^{-\kappa\tau}}{\kappa},
\]
and for SABR, in $x=\log(F/K)$,
\[
-\calM[h]=2\rho\nu\alpha W\,\partial_{xW}h+\nu^2W\,\partial_Wh+2\nu^2W^2\,\partial_{WW}h,
\]
where $\partial_W$ denotes the derivative of the Black--Scholes price with
respect to the integrated variance at fixed $x$ and $\tau$. These
derivatives are explicit, and the right-hand sides stay bounded as
$\tau\to0$. The dependence of $h$ on the variance matters: subtracting a
Black--Scholes price with one fixed volatility removes the kink from the
payoff, but leaves a right-hand side that grows like $\tau^{-1/2}$ near the
strike. For SABR, the far-field condition on the faces in $\alpha$
(Section~\ref{sec:sabr-exp}) is imposed on $w$ instead of on $\Vb$. Unlike
for the jump models, the right-hand side now depends on all the parameters
of the operator, through $h$ and $W$, so the exact Jacobian of
Section~\ref{sec:calibration} would also need the derivatives of the
right-hand side. We therefore use this subtraction only for pricing
(Sections~\ref{sec:heston-exp} and~\ref{sec:sabr-exp}).}

\section{Sensitivities and calibration}\label{sec:calibration}

The fitted price is an explicit function of the state variables and,
through the output weights, of the model parameters. Differentiating it
with respect to the state gives the Greeks (Section~\ref{sec:greeks}).
Differentiating it with respect to the model parameters gives the
Jacobian needed for calibration (Sections~\ref{sec:implicit}
to~\ref{sec:v0}). Section~\ref{sec:calib-procedure} describes the
calibration procedure, and Section~\ref{sec:identifiability} discusses
which parameters a calibration can recover.

\subsection{Greeks}\label{sec:greeks}

Write the fitted normalised price as
$u(z)=\sum_{k}y_k^\ast\,\Phi(s_k(z))$, so that the price of the call in
currency units is $\widehat V(S,t)=K\,u\big(\log(S/K),\,T-t\big)$. The
derivatives of $u$ are sums of the feature derivatives
\eqref{eq:feature-derivs} with the same weights, for example
\[
u_x=\sum_k y_k^\ast A_k^{(x)}\phi(s_k),\qquad
u_{xx}=-\sum_k y_k^\ast \big(A_k^{(x)}\big)^2 s_k\,\phi(s_k),\qquad
u_\tau=\sum_k y_k^\ast A_k^{(\tau)}\phi(s_k).
\]
Changing variables from $(x,\tau)$ back to $(S,t)$ gives
\begin{equation*}
\Delta=\frac{\partial\widehat V}{\partial S}=\frac{K}{S}\,u_x,
\qquad
\Gamma=\frac{\partial^2\widehat V}{\partial S^2}=\frac{K}{S^2}\big(u_{xx}-u_x\big),
\qquad
\Theta=\frac{\partial\widehat V}{\partial t}=-K\,u_\tau .
\end{equation*}
For Heston, the sensitivity to the current variance is
$\partial\widehat V/\partial v=K\sum_k y_k^\ast A_k^{(v)}\phi(s_k)$, and
similarly for the SABR volatility. Each Greek is one further evaluation
of the features at the weights already computed; no new solve is needed.
When the Black--Scholes price is subtracted, the Greeks of
$u_{\mathrm{BS}}$, which are explicit, are added to those of the fitted
remainder. When results are averaged over several feature draws, the
Greeks are averaged in the same way as the prices.

\subsection{Parameter sensitivities}\label{sec:implicit}

Calibration needs the derivative of the fitted price with respect to each
model parameter $\theta_p$. Unlike the state, a model parameter does not
enter the features: it enters the least-squares problem, because it
changes the pricing operator and hence the matrix $\mathbf A$ and, when
the Black--Scholes price is subtracted, the vector $\mathbf b$. The output
weights $\mathbf y^\ast$ therefore depend on $\theta_p$, and their
derivative is obtained by differentiating the normal equations.

\begin{proposition}[Parameter sensitivity of the output weights]\label{prop:sens}
Let $\mathbf H=\mathbf A^\top\mathbf A+\reg I$ and
$\mathbf e=\mathbf b-\mathbf A\mathbf y^\ast$ be the matrix and the
residual of the solve \eqref{eq:lsq-reg}, and let
$\mathbf A_p=\partial\mathbf A/\partial\theta_p$ and
$\mathbf b_p=\partial\mathbf b/\partial\theta_p$. Then
\begin{equation}\label{eq:sens}
\frac{\partial\mathbf y^\ast}{\partial\theta_p}
 = \mathbf H^{-1}\Big(\mathbf A_p^\top\mathbf e
   - \mathbf A^\top\mathbf A_p\,\mathbf y^\ast
   + \mathbf A^\top\mathbf b_p\Big).
\end{equation}
\end{proposition}

\begin{proof}
Differentiate $\mathbf H\mathbf y^\ast=\mathbf A^\top\mathbf b$ with
respect to $\theta_p$. Since
$\partial\mathbf H/\partial\theta_p=\mathbf A_p^\top\mathbf A+\mathbf A^\top\mathbf A_p$,
\[
\mathbf H\,\frac{\partial\mathbf y^\ast}{\partial\theta_p}
=\mathbf A_p^\top\mathbf b+\mathbf A^\top\mathbf b_p
-\big(\mathbf A_p^\top\mathbf A+\mathbf A^\top\mathbf A_p\big)\mathbf y^\ast ,
\]
and $\mathbf A_p^\top\mathbf b-\mathbf A_p^\top\mathbf A\mathbf y^\ast
=\mathbf A_p^\top\mathbf e$.
\end{proof}

Formula \eqref{eq:sens} is the classical derivative of a regularised
least-squares solution with respect to parameters on which the problem
depends \citep{golub1973}. Three features make it cheap and exact here.

\begin{itemize}[leftmargin=*]
\item \emph{The matrix is already factorised.} The matrix $\mathbf H$ in
  \eqref{eq:sens} is the same matrix whose Cholesky factorisation
  $\mathbf H=\mathbf L\mathbf L^\top$, with $\mathbf L$ lower triangular,
  was computed for the price. Solving \eqref{eq:sens} for one parameter
  therefore needs only two triangular solves with $\mathbf L$ and
  $\mathbf L^\top$, at a cost of $O(\nfeat^2)$ operations, instead of a
  new factorisation, which would cost $O(\nfeat^3)$. Forming the
  right-hand side costs $O(N\nfeat)$ operations.
\item \emph{Only the interior rows change.} The payoff and the boundary
  data do not depend on the model parameters
  (Section~\ref{sec:boundary}), so the initial and boundary rows of
  $\mathbf A$ and $\mathbf b$ are the same for every $\theta$. The rows of
  $\mathbf A_p$ for the interior points are the parameter derivatives of
  the explicit expressions of Lemmas~\ref{lem:pointwise}
  and~\ref{lem:jump}, and are themselves explicit; they are listed in
  Appendix~\ref{app:param-derivs}. In the plain solve $\mathbf b_p=0$.
  When the Black--Scholes price is subtracted, the interior entries of
  $\mathbf b$ are the values of the right-hand side $\calF$ of
  \eqref{eq:bs-source}, and $\mathbf b_p$ is their explicit derivative.
\item \emph{The result is exact.} The derivative of the fitted price at a
  quoted point $z$ is
  \begin{equation}\label{eq:price-sens}
  \frac{\partial\widehat V(z)}{\partial\theta_p}
  =K\sum_k \Phi(s_k(z))\,\frac{\partial y_k^\ast}{\partial\theta_p},
  \end{equation}
  plus, when the Black--Scholes price is subtracted, $K\,\partial
  u_{\mathrm{BS}}/\partial\theta_p$ (the Black--Scholes vega for
  $\theta_p=\sigma$, and zero otherwise). This is the exact derivative of
  the solver's prices; no prices are differenced.
\end{itemize}

\subsection{Parameter-separable operators}\label{sec:separable}

\chg{For the Heston and SABR models the model parameters enter the
interior rows only through a few scalar coefficients, and this makes
repeated solves at different parameter values much cheaper.} From
\eqref{eq:PkQk-heston}, the Heston operator applied
to a feature can be written as
\begin{equation*}
\calM[\Phi(s_k)]=\sum_{i=0}^{4} c_i(\theta)\,\mathcal D_i[\Phi(s_k)],
\qquad
c(\theta)=\big(1,\ -\kappa\vbar,\ \kappa,\ \rho\xi,\ \tfrac12\xi^2\big),
\end{equation*}
where the terms
\begin{equation*}
	\left\{\begin{array}{l}
\mathcal D_0[\Phi(s_k)]=r\,\Phi(s_k)+\Big(\tfrac12 v\big(A_k^{(x)}\big)^2s_k
  +A_k^{(\tau)}-\big(r-q-\tfrac v2\big)A_k^{(x)}\Big)\phi(s_k),\\ [6pt]
\mathcal D_1[\Phi(s_k)]=A_k^{(v)}\phi(s_k),\qquad
\mathcal D_2[\Phi(s_k)]=v\,A_k^{(v)}\phi(s_k),\\ [6pt]
\mathcal D_3[\Phi(s_k)]=v\,A_k^{(x)}A_k^{(v)}s_k\,\phi(s_k),\qquad
\mathcal D_4[\Phi(s_k)]=v\,\big(A_k^{(v)}\big)^2 s_k\,\phi(s_k)
\end{array}\right.
\end{equation*} 
do not depend on the model parameters. For SABR, similarly,
$c(\theta)=(1,\ \rho\nu,\ \tfrac12\nu^2)$, with
$\mathcal D_0=(\tfrac12\alpha^2F^{2\beta}(A_k^{(F)})^2s_k+A_k^{(\tau)})\phi(s_k)$,
$\mathcal D_1=\alpha^{2} F^\beta A_k^{(F)}A_k^{(\alpha)}s_k\,\phi(s_k)$ and
$\mathcal D_2=\alpha^2(A_k^{(\alpha)})^2s_k\,\phi(s_k)$.

Writing $\mathbf D_i$ for the matrix with entries
$\mathcal D_i[\Phi(s_k)](z_j)$ over the interior points, the interior
block of $\mathbf A$ is $\sum_i c_i(\theta)\mathbf D_i$, and
\[
\mathbf A^\top\mathbf A=\sum_{i,j}c_i(\theta)\,c_j(\theta)\,\mathbf D_i^\top\mathbf D_j
+\mathbf B^\top\mathbf B ,
\]
where $\mathbf B$ collects the initial and boundary rows. The products
$\mathbf D_i^\top\mathbf D_j$ and $\mathbf B^\top\mathbf B$ do not depend
on $\theta$ and are computed once, at a cost of $O(N\nfeat^2)$ each. A
solve at a new parameter vector then needs only the weighted sum of these
$\nfeat\times\nfeat$ matrices and one Cholesky factorisation, at a cost
independent of the number $N$ of collocation points. The derivatives
$\mathbf A_p=\sum_i(\partial c_i/\partial\theta_p)\mathbf D_i$ enter the
right-hand side of \eqref{eq:sens} in the same way. This is a standard
splitting into a parameter-independent precomputation and a cheap
parameter-dependent step. \chg{For the jump models, the parameters of the
jump law ($\mu_J,\sigma_J$ for Merton; $p,\eta_1,\eta_2$ for Kou) enter the
jump term nonlinearly, through $\Phi$ and $\operatorname{erfcx}$, so the
interior rows are reassembled at each trial parameter vector. This costs
$O(N\nfeat^2)$ operations per solve, which is still a matter of seconds
for the sizes used here.}

\subsection{Parameters that only move the evaluation point}\label{sec:v0}

The current variance $v_0$ in Heston and the current volatility
$\alpha_0$ in SABR are calibrated as well, but they do not appear in the
pricing equation: they only fix the point at which the fitted surface is
evaluated. Changing them requires no new solve, and their sensitivity is a
state derivative in the sense of Section~\ref{sec:greeks},
\begin{equation}\label{eq:v0-sens}
\frac{\partial\widehat V}{\partial v_0}
 = K\sum_k y_k^\ast\,A_k^{(v)}\,\phi(s_k),
\end{equation}
with $s_k$ evaluated at the quoted point $(x,v_0,\tau)$, and similarly for
$\alpha_0$ with $A_k^{(\alpha)}$.

\subsection{The calibration procedure}\label{sec:calib-procedure}

Let $V_i^{\mathrm{mkt}}$, $i=1,\dots,n_q$, be quoted prices of options
with log-moneyness $x_i$ and maturity $\tau_i$, and let
$\widehat V_i(\theta)$ be the corresponding prices of the solver at the
parameter vector $\theta$. All of them come from a single solve: in the
normalised variables of Section~\ref{sec:setting} every call has the same
payoff and the models are time-homogeneous, so one solve on
$\tau\in[0,\max_i\tau_i]$ gives the whole surface, and each quote is an
evaluation of it, $\widehat V_i(\theta)=K_i\sum_k y_k^\ast(\theta)\,
\Phi(s_k(x_i,\tau_i))$ (for Heston and SABR at the current variance or
volatility). We choose $\theta$ to minimise the sum of
squared pricing errors over a box $\Theta$ of admissible parameter values,
\begin{equation*}
\widehat\theta=\argmin_{\theta\in\Theta}\ \dfrac12\sum_{i=1}^{n_q}
\big(\widehat V_i(\theta)-V_i^{\mathrm{mkt}}\big)^2 .
\end{equation*}
This is a nonlinear least-squares problem in the $\npar$ parameters,
solved iteratively. One iteration at the current trial vector $\theta$
consists of three steps:
\begin{enumerate}[leftmargin=*]
\item solve the pricing problem at $\theta$, which gives the weights
  $\mathbf y^\ast(\theta)$ and the Cholesky factor of $\mathbf H$, and
  evaluate the residuals $\widehat V_i(\theta)-V_i^{\mathrm{mkt}}$;
\item compute the Jacobian
  $J_{ip}=\partial\widehat V_i/\partial\theta_p$ from
  Proposition~\ref{prop:sens} and \eqref{eq:price-sens}, or
  from \eqref{eq:v0-sens} for $v_0$ and $\alpha_0$, reusing the factor;
\item update $\theta$ with a derivative-based least-squares step.
\end{enumerate}
\chg{We use the trust-region reflective method of \citet{branch1999}, as
implemented in SciPy, which handles the bounds of $\Theta$ directly;
other derivative-based least-squares methods, such as
Levenberg--Marquardt, can be used in the same way.} The features and the
collocation points are drawn once, before the calibration starts, and kept
fixed throughout, so the solver's prices and their Jacobian are
deterministic functions of $\theta$. When several feature draws are used,
the prices and the Jacobians are averaged over the draws.

\subsection{Can the parameters be recovered?}\label{sec:identifiability}

\chgg{A calibrated model is used to price and hedge other products, so its
parameters must be right, not only its fit to the quotes. A calibration
can fit the quotes well and still return wrong parameters, for two
reasons.

The first is the model. Different parameter vectors may give almost the
same prices at the quoted points: many small jumps look much like extra
diffusion; in Heston, the mean-reversion speed $\kappa$ has little effect
on options of a year or less, and that effect is largely offset by $\xi$
and $\rho$ (Section~\ref{sec:heston-exp}); in SABR, $\beta$ and $\rho$
have similar effects on the smile, which is why $\beta$ is usually fixed
\citep{hagan2002}. No solver can remove this difficulty. The Jacobian $J$
of Section~\ref{sec:calib-procedure}, $J_{ip}=\partial\widehat
V_i/\partial\theta_p$, measures it: a parameter that hardly moves the
quotes, or two parameters that move them in almost the same way, show up
as a small singular value of $J$, that is, a direction in parameter space
along which the quotes barely change. Singular values are meaningful only
when the parameters are measured in comparable units, for example as
relative changes.

The second is the solver. Let $V(\theta)$ be the exact prices at the
quoted points, suppose the quotes are exact prices at a true vector
$\theta^\star$, and let $\varepsilon$ bound the solver's pricing error,
$\|\widehat V(\theta)-V(\theta)\|\le\varepsilon$, in the Euclidean norm
over the quotes. The calibrated vector $\widehat\theta$ fits the quotes at
least as well as $\theta^\star$, so
$\|\widehat V(\widehat\theta)-V(\theta^\star)\|\le\varepsilon$, and by the
triangle inequality $\|V(\widehat\theta)-V(\theta^\star)\|\le2\varepsilon$:
any parameter vector whose exact prices lie within $2\varepsilon$ of the
quotes may be returned. To first order,
$V(\widehat\theta)-V(\theta^\star)\approx J\,(\widehat\theta-\theta^\star)$,
so
\begin{equation}\label{eq:param-error}
\|\widehat\theta-\theta^\star\|\ \lesssim\ \frac{2\varepsilon}{\sigma_{\min}(J)},
\end{equation}
where $\sigma_{\min}(J)$ is the smallest singular value of $J$. The
parameter error can be large when the model identifies some parameters
poorly (small $\sigma_{\min}$), when the solver is not accurate enough
(large $\varepsilon$), and most of all when both happen. A small solver
error therefore guarantees that the parameters are recovered, and the
weaker the identification, the smaller the error must be. The bound is a
worst case: an error does most harm when it resembles the effect of a
change in the poorly determined parameters. If the quotes are generated
by the solver itself, $\varepsilon$ plays no role and the parameters are
recovered whenever $\sigma_{\min}(J)>0$.

Both quantities can be assessed without a reference price. The Jacobian
at $\widehat\theta$ is computed anyway in the last step of the
calibration, and its singular vectors show which directions in parameter
space the quotes determine poorly. The solver's error can be gauged by
repeating the solve with independent feature draws and measuring how
much the prices at the quoted points change; this captures only the part
of the error that varies between draws, so it is an indication rather
than a bound. More simply, if the calibration repeated with independent
draws returns parameters that differ by more than the accuracy required,
the solver is not accurate enough for these quotes.
Section~\ref{sec:jump-calib} shows these situations and checks for the
jump models.}

\section{Numerical experiments}\label{sec:experiments}

The experiments test the claims of Section~\ref{sec:contributions} on
problems for which accurate reference prices exist.
Section~\ref{sec:merton-exp} prices the two jump models and compares the
result with finite differences. Section~\ref{sec:convergence}
\chgg{compares the two methods at equal run time}.
Section~\ref{sec:jump-calib} calibrates the jump models and shows the role
of the solver's accuracy in recovering the parameters.
Section~\ref{sec:bansal} compares the method with a physics-informed
neural network on that network's own test problems.
Sections~\ref{sec:heston-exp} and~\ref{sec:sabr-exp} apply the method,
\chgo{mainly with the plain solve}, to Heston and SABR\chgo{, with a test of
the subtraction for each}.

Table~\ref{tab:setup} lists the model parameters and
the main solver settings. All other settings, such as the bandwidths
\chgg{and the placement of the collocation points}, are given in the
accompanying code \chgo{(see Code availability)}. Errors are
measured against references computed independently of the solver: the
exact series for Merton \citep{merton1976}, the COS method for Kou
\citep{fang2008}, Fourier integration for Heston, and Hagan's formula and
Monte Carlo simulation for SABR. Reported prices are averages over
$\nens$ independent feature draws. \chg{Table~\ref{tab:setup} also gives the domain on which each problem is
solved; all errors are measured inside it, away from its faces.}

\chgg{For the jump models, every result comes from one script, with one
setting for pricing and one for calibration, and every pricing error is
measured on the same grid $G$: $81$ equally spaced points in
$x\in[-0.8,0.8]$ and $32$ maturities, eight equally spaced in each of the
bands $[0.005,0.05]$, $[0.05,0.2]$, $[0.2,0.5]$ and $[0.5,1]$. We report
the root-mean-square error (RMSE) over $G$ of the price divided by the
strike. Both references were checked against more accurate versions of
themselves (a longer series; a COS expansion with twice as many terms on a
wider interval), and agree with them to within $3\times10^{-10}$ on $G$,
far below every error reported. The loss weights are
$w_{\mathrm{ic}}=\sqrt{40}$ and $w_{\mathrm b}=5$, with $7000$ (Merton)
or $7500$ (Kou) interior points, $1600$ initial points and $300$ boundary
points; for Heston and SABR they are $w_{\mathrm{ic}}=10$ and
$w_{\mathrm b}=1$, with $8500$ interior, $900$ initial and $600$ boundary
points in $x$ (SABR has $200$ more on the faces in $\alpha$). All
computations run on one machine with an Intel Xeon Gold 6138 processor
($2.0$\,GHz), and each jump-model run uses one thread.}

\begin{table}[htbp]
\centering
\caption{Model parameters, \chg{solution domain} and main solver settings.
\chg{$x=\log(S/K)$ for the
jump models and Heston, and $x=\log(F/K)$ for SABR. For Heston and SABR,
each feature draw is itself an average of three (Heston) or five (SABR)
networks; $8\times3$ means eight such draws.} \chgg{The jump-model
settings hold for the plain solve and for the solve with subtraction; one
third of the features are placed near the strike
(Section~\ref{sec:bs-subtraction}).}}
\label{tab:setup}
\small
\begin{tabular}{>{\raggedright}p{0.13\textwidth}>{\raggedright}p{0.27\textwidth}>{\raggedright}p{0.2\textwidth}>{\raggedright}p{0.11\textwidth}>{\raggedright\arraybackslash}p{0.12\textwidth}}
\toprule
Experiment & Model parameters & \leavevmode\chg{Domain} & $\nfeat$ & $\nens$ \\
\midrule
Merton pricing & $r=0.05$, $T=1$, $\sigma=0.2$, $\lambda=0.1$, $\mu_J=-0.1$, $\sigma_J=0.3$ & \leavevmode\chg{$x\in[-1.2,1.2]$, $\tau\in[0,1]$}\chgg{; finite differences on $[-2.5,2.5]$} & \leavevmode\chgg{$1800$} & \leavevmode\chgg{$4$} \\
Kou pricing & $r=0.05$, $T=1$, $\sigma=0.2$, $\lambda=0.3$, $p=0.4$, $\eta_1=10$, $\eta_2=5$ & \leavevmode\chg{$x\in[-1.2,1.2]$, $\tau\in[0,1]$}\chgg{; finite differences on $[-2.5,2.5]$} & \leavevmode\chgg{$1800$} & \leavevmode\chgg{$4$} \\
Jump calibration & as for pricing & \leavevmode\chg{$x\in[-1.2,1.2]$, $\tau\in[0,1]$} & \leavevmode\chgg{$800$} & \leavevmode\chgg{$1$ ($5$ in the repeat study)} \\
\leavevmode\chg{PINN comparison} & \leavevmode\chg{as in \citet{bansal2026}} & \leavevmode\chg{$x\in[-1,1]$, $\tau\in[0,1]$} & \leavevmode\chg{$800$, $400$, $300$} & \leavevmode\chg{$5$ (median)} \\
Heston & Cases 1--4 of \citet{inthout2010} (Table~\ref{tab:heston-adi}) & \leavevmode\chg{$x\in[\log 0.1,\log 8]$, $v\in[0.005,0.25]$, $\tau\in[0,3.5]$ (pricing), $[0,1.5]$ (calibration)} & \leavevmode\chg{$8000$} & \leavevmode\chg{$8\times3$ (pricing), $3$ (calibration)} \\
SABR & $\beta=1$, \chg{$r=0.02$,} $\rho=-0.3$, $\nu=0.4$, $\alpha_0=0.25$ & \leavevmode\chg{$x\in[\log 0.1,\log 8]$, $\alpha\in[0.005,0.55]$, $\tau\in[0,2]$ (pricing), $[0,2.5]$ (calibration), $[0,6]$ (Monte Carlo comparison)} & \leavevmode\chg{$4000$} & \leavevmode\chg{$5\times5$ (pricing), $5$ (otherwise)} \\
\bottomrule
\end{tabular}
\end{table}

Unless stated otherwise, we use asymptotic \chgb{boundary}
conditions on the faces in $x$. Far out of the money the call is almost
worthless, and far in the money it behaves like a forward contract, so
\begin{equation}\label{eq:bc-asymptotic}
\Vb(x_{\min},\tau) = 0,
\qquad
\Vb(x_{\max},\tau) = e^{x_{\max}-q\tau} - e^{-r\tau}.
\end{equation}
\chgg{With the Black--Scholes subtraction we set $w=0$ on the faces in
$x$ instead, since the Black--Scholes price already behaves in this way
far from the strike.} These conditions do not depend on the model
parameters and require no reference prices.

\subsection{Pricing the jump models}\label{sec:merton-exp}\label{sec:kou-exp}

\chgg{This study shows four things: the exact jump term prices both models
without quadrature; the payoff kink limits the accuracy of the plain
solve; the Black--Scholes subtraction removes that limit; and the
resulting accuracy is close to that of a standard finite-difference
scheme, though far from that of a finite-difference scheme that uses the
subtraction as well.

All solvers use the boundary conditions above, so no reference price
enters any of them. The domain $x\in[-1.2,1.2]$ is wide enough: replacing
these boundary values by the exact ones changes the prices of the solve
with subtraction on $G$ by an RMSE of $2\times10^{-8}$ for Merton and
$1.4\times10^{-7}$ for Kou, a small fraction of its error. A wider domain
does not help, because the same number of features is then spread over a
larger region: from $x\in[-1.2,1.2]$ to $[-2.5,2.5]$ the Merton error
grows tenfold, and the Kou error hardly changes. The finite-difference
scheme runs on $[-2.5,2.5]$, where its truncation error is also
negligible.

With the plain solve, the relative errors at $\tau=1$ over
$S/K\in[0.7,1.3]$ range from $0.005\%$ to $0.5\%$ for Merton and from
$0.003\%$ to $0.04\%$ for Kou. They are largest at the deepest
out-of-the-money strike, where the price itself is smallest. With the
Black--Scholes subtraction they are below $0.003\%$ for Merton and
$0.015\%$ for Kou at every strike.

Table~\ref{tab:jump-fdm} compares the two solves with finite
differences. The scheme uses Crank--Nicolson time stepping with Rannacher
start-up (four implicit Euler half-steps), $4001$ grid points and $2000$
time steps. Its jump integral is computed at every step by
Gauss--Hermite (Merton, $64$ nodes) or Gauss--Laguerre (Kou, $48$ nodes)
quadrature applied to the grid values, interpolated between grid points by
cubic polynomials. We report three versions: with the jump term taken
from the previous time step (\emph{explicit}), which is first-order
accurate in time; with the jump term treated implicitly by a fixed-point
iteration at every step \citep{dhalluin2005}, which is the standard
choice; and the implicit version applied to the remainder
$w=\Vb-u_{\mathrm{BS}}$, with exactly the same Black--Scholes subtraction
as in Section~\ref{sec:bs-subtraction}.

The implicit scheme without subtraction has an overall error of
$5\times10^{-7}$ for both models. \PIRanNN{} with subtraction is as
accurate for Merton ($4.5\times10^{-7}$) and about four times less
accurate for Kou ($1.9\times10^{-6}$). For Merton it is more accurate at
the shortest maturities, where the kink affects the finite-difference
scheme most, and less accurate at long maturities. Treating the jump term
explicitly raises the finite-difference error at long maturities, to
$7.2\times10^{-7}$ (Merton) and $1.7\times10^{-6}$ (Kou) overall. With
the subtraction, the finite-difference error falls to
$2.5\times10^{-9}$ and $7.1\times10^{-9}$, about $180$ and $270$ times
below \PIRanNN{} with subtraction. The subtraction is thus just as useful
to a finite-difference scheme, and in one space dimension a
finite-difference scheme that uses it is the more accurate pricer.

Two further details of the finite-difference scheme matter at this level
of accuracy, and neither arises for \PIRanNN{}, whose price is a formula
in $x$ and $\tau$. A finite-difference price is available only at the
grid points and time steps. Taking each price of $G$ at the nearest time
step instead of the exact maturity raised the finite-difference error in
our tests about fivefold; interpolating linearly instead of by cubic
splines between grid points changed it by less than $25\%$.}

\begin{table}[htbp]
\centering
\caption{\chgg{Price RMSE on the grid $G$, by maturity band. \PIRanNN{}
uses $1800$ features and averages four feature draws. The
finite-difference scheme (FDM) uses Crank--Nicolson time stepping with
Rannacher start-up, $4001$ grid points on $[-2.5,2.5]$ and $2000$ time
steps; ``explicit'' and ``implicit'' refer to its treatment of the jump
term.}}
\label{tab:jump-fdm}
\small
\begin{tabular}{lccccc}
\toprule
Maturity $\tau$ & $[0.005,0.05]$ & $[0.05,0.2]$ & $[0.2,0.5]$ & $[0.5,1]$ & overall \\
\midrule
\multicolumn{6}{l}{\emph{Merton (exact reference)}} \\
\chgg{\PIRanNN{}, plain}              & \chgg{$5.9\times10^{-5}$} & \chgg{$1.2\times10^{-5}$} & \chgg{$2.8\times10^{-5}$} & \chgg{$5.1\times10^{-5}$} & \chgg{$4.3\times10^{-5}$} \\
\chgg{\PIRanNN{}, with subtraction}   & \chgg{$3.9\times10^{-7}$} & \chgg{$1.1\times10^{-7}$} & \chgg{$3.7\times10^{-7}$} & \chgg{$6.6\times10^{-7}$} & \chgg{$4.5\times10^{-7}$} \\
\chgg{FDM, explicit jump term}        & \chgg{$9.3\times10^{-7}$} & \chgg{$3.7\times10^{-7}$} & \chgg{$5.0\times10^{-7}$} & \chgg{$8.7\times10^{-7}$} & \chgg{$7.2\times10^{-7}$} \\
\chgg{FDM, implicit jump term}        & \chgg{$9.3\times10^{-7}$} & \chgg{$3.0\times10^{-7}$} & \chgg{$2.0\times10^{-7}$} & \chgg{$1.5\times10^{-7}$} & \chgg{$5.0\times10^{-7}$} \\
\chgg{FDM, implicit, with subtraction} & \chgg{$4.3\times10^{-9}$} & \chgg{$1.3\times10^{-9}$} & \chgg{$1.1\times10^{-9}$} & \chgg{$1.8\times10^{-9}$} & \chgg{$2.5\times10^{-9}$} \\
\midrule
\multicolumn{6}{l}{\emph{Kou (COS reference)}} \\
\chgg{\PIRanNN{}, plain}              & \chgg{$6.8\times10^{-5}$} & \chgg{$2.7\times10^{-5}$} & \chgg{$1.8\times10^{-5}$} & \chgg{$3.7\times10^{-5}$} & \chgg{$4.2\times10^{-5}$} \\
\chgg{\PIRanNN{}, with subtraction}   & \chgg{$3.3\times10^{-6}$} & \chgg{$1.7\times10^{-6}$} & \chgg{$1.1\times10^{-6}$} & \chgg{$9.6\times10^{-7}$} & \chgg{$1.9\times10^{-6}$} \\
\chgg{FDM, explicit jump term}        & \chgg{$9.8\times10^{-7}$} & \chgg{$6.7\times10^{-7}$} & \chgg{$1.4\times10^{-6}$} & \chgg{$2.7\times10^{-6}$} & \chgg{$1.7\times10^{-6}$} \\
\chgg{FDM, implicit jump term}        & \chgg{$9.6\times10^{-7}$} & \chgg{$3.1\times10^{-7}$} & \chgg{$2.0\times10^{-7}$} & \chgg{$1.5\times10^{-7}$} & \chgg{$5.2\times10^{-7}$} \\
\chgg{FDM, implicit, with subtraction} & \chgg{$1.3\times10^{-8}$} & \chgg{$4.0\times10^{-9}$} & \chgg{$2.8\times10^{-9}$} & \chgg{$4.1\times10^{-9}$} & \chgg{$7.1\times10^{-9}$} \\
\bottomrule
\end{tabular}
\end{table}

\chgb{To see where the gain comes from,} Table~\ref{tab:ablation} separates
the two ingredients of Section~\ref{sec:bs-subtraction}, with the total
number of features held fixed. \chgg{For Merton, the extra features near
the strike alone reduce the error by a factor of about $7$, the
subtraction alone by about $100$, and the two together by about $700$;
for Kou the factors are about $15$, $70$ and $300$.} The subtraction is the main source of the gain.
Appendix~\ref{app:robustness}
repeats the Merton study over a range of jump parameters\chgg{, with an
earlier version of the solver}: the plain solve stays near $10^{-4}$
throughout, and the
subtraction gains about two orders of magnitude for every parameter set.

\begin{table}[htbp]
\centering
\caption{Price RMSE on $G$ with and without the two ingredients of the
Black--Scholes subtraction step. \chgg{Every run uses $1800$ features and
averages four feature draws, as in Table~\ref{tab:jump-fdm}.}}
\label{tab:ablation}
\begin{tabular}{lcc}
\toprule
Configuration & Merton & Kou \\
\midrule
\chg{neither ingredient}                 & \chgg{$3.2\times10^{-4}$} & \chgg{$6.1\times10^{-4}$} \\
$+$ extra features near the strike       & \chgg{$4.3\times10^{-5}$} & \chgg{$4.2\times10^{-5}$} \\
$+$ Black--Scholes subtraction           & \chgg{$3.1\times10^{-6}$} & \chgg{$9.2\times10^{-6}$} \\
$+$ both                                 & \chgg{$4.5\times10^{-7}$} & \chgg{$1.9\times10^{-6}$} \\
\bottomrule
\end{tabular}
\end{table}

\chg{To isolate the effect
of the exact jump term, we repeat the \chgg{solve with subtraction} with the jump term of
each feature computed instead by Gauss--Hermite (Merton) or
Gauss--Laguerre (Kou) quadrature with $Q$ nodes, everything else
unchanged (Table~\ref{tab:quad}). \chgg{We use the solve with subtraction
because its small error tests the jump term more strictly than the plain
solve would.} The error against the reference
includes the error of the solve itself, which is the same for every
variant, so the table reports each variant's error relative to that of the
exact term. \chgg{Quadrature reaches the accuracy of the exact term with
about $12$ nodes for Merton and $8$ for Kou. For Merton the exact term is
then about $12$ times cheaper to assemble; for Kou, where it requires the
function $\operatorname{erfcx}$, it costs about as much as five nodes.} The
exact term also removes the choice of $Q$.}

\begin{table}[htbp]
\centering
\caption{\chgg{Solve with subtraction, with the jump term of each feature
computed exactly or by $Q$-node quadrature: price RMSE on $G$ relative to
the RMSE with the exact term, and time to assemble the jump term, summed
over the four feature draws. The RMSE with the exact term is
$4.5\times10^{-7}$ for Merton and $1.9\times10^{-6}$ for Kou.}}
\label{tab:quad}
\begin{tabular}{lcc@{\hspace{1cm}}lcc}
\toprule
\multicolumn{3}{c}{Merton} & \multicolumn{3}{c}{Kou} \\
\cmidrule(lr){1-3}\cmidrule(lr){4-6}
Jump term & rel.\ RMSE & time (s) & Jump term & rel.\ RMSE & time (s) \\
\midrule
exact      & $1$     & \chgg{$1.6$}  & exact      & $1$     & \chgg{$18.0$} \\
$Q=2$      & \chgg{$17.7$}  & \chgg{$3.4$}  & $Q=2$      & \chgg{$5.99$}  & \chgg{$7.8$} \\
$Q=4$      & \chgg{$4.04$}  & \chgg{$6.7$}  & $Q=4$      & \chgg{$1.65$}  & \chgg{$14.8$} \\
$Q=8$      & \chgg{$1.11$}  & \chgg{$13.0$} & $Q=8$      & \chgg{$1.01$}  & \chgg{$28.8$} \\
$Q=12$     & \chgg{$1.01$}  & \chgg{$19.6$} & \chgg{$Q=16$}     & \chgg{$1.00$}  & \chgg{$55.7$} \\
\chgg{$Q=20$}     & \chgg{$1.00$}  & \chgg{$32.8$} & $Q=32$     & \chgg{$1.00$}  & \chgg{$108.5$} \\
$Q=40$     & \chgg{$1.00$}  & \chgg{$65.5$} &  &  &  \\
\bottomrule
\end{tabular}
\end{table}

The Greeks of Section~\ref{sec:greeks} come from the
same solve. \chgg{Table~\ref{tab:greeks} compares Delta and Gamma of the
solve with subtraction with the reference values, at a short and a long
maturity. Delta is within $0.07\%$ (Merton) and $0.3\%$ (Kou) of the
reference, and Gamma within $0.3\%$ and $0.7\%$. The largest Delta
errors, and Kou's largest Gamma errors, occur at $\tau=0.1$ away from the
strike, where the Greeks themselves are small.}

\begin{table}[htbp]
\centering
\caption{\chgg{Relative error (in $\%$) of Delta and Gamma from the solve
with subtraction. Reference Greeks are central differences of the
reference price.}}
\label{tab:greeks}
\small
\begin{tabular}{ccccc@{\hspace{6mm}}cccc}
\toprule
 & \multicolumn{4}{c}{Merton} & \multicolumn{4}{c}{Kou} \\
\cmidrule(lr){2-5}\cmidrule(lr){6-9}
 & \multicolumn{2}{c}{$\tau=0.1$} & \multicolumn{2}{c}{$\tau=1$} & \multicolumn{2}{c}{$\tau=0.1$} & \multicolumn{2}{c}{$\tau=1$} \\
$S/K$ & $\Delta$ & $\Gamma$ & $\Delta$ & $\Gamma$ & $\Delta$ & $\Gamma$ & $\Delta$ & $\Gamma$ \\
\midrule
\chgg{$0.85$} & \chgg{$0.072$} & \chgg{$0.11$}  & \chgg{$0.035$} & \chgg{$0.25$}  & \chgg{$0.27$}  & \chgg{$0.58$}  & \chgg{$0.002$} & \chgg{$0.074$} \\
\chgg{$0.95$} & \chgg{$0.005$} & \chgg{$0.013$} & \chgg{$0.002$} & \chgg{$0.28$}  & \chgg{$0.018$} & \chgg{$0.089$} & \chgg{$0.002$} & \chgg{$0.043$} \\
\chgg{$1.00$} & \chgg{$0.002$} & \chgg{$0.010$} & \chgg{$0.010$} & \chgg{$0.20$}  & \chgg{$0.009$} & \chgg{$0.007$} & \chgg{$<0.001$} & \chgg{$0.077$} \\
\chgg{$1.05$} & \chgg{$0.002$} & \chgg{$0.017$} & \chgg{$0.002$} & \chgg{$0.009$} & \chgg{$0.009$} & \chgg{$0.066$} & \chgg{$<0.001$} & \chgg{$0.025$} \\
\chgg{$1.15$} & \chgg{$<0.001$} & \chgg{$0.065$} & \chgg{$0.012$} & \chgg{$0.11$}  & \chgg{$0.003$} & \chgg{$0.67$}  & \chgg{$<0.001$} & \chgg{$0.17$} \\
\bottomrule
\end{tabular}
\end{table}

\subsection{\chgg{Accuracy and run time}}\label{sec:convergence}

\begin{figure}[htbp]
\centering
\includegraphics[width=\textwidth]{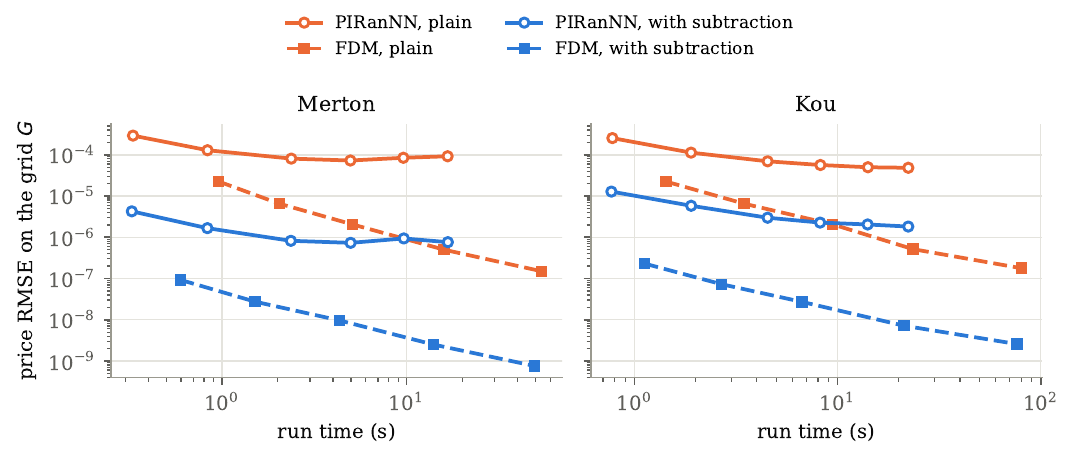}
\caption{\chgg{Price RMSE on $G$ against run time, for one feature draw of
\PIRanNN{} with $300$ to $3000$ features (circles; mean over four single
draws) and for the Crank--Nicolson scheme with implicit jump term with
$501$ to $8001$ grid points (squares), each plain and with
the Black--Scholes subtraction. The collocation points of the network are
the same for every $\nfeat$.}}
\label{fig:convergence}
\end{figure}

\chgg{Figure~\ref{fig:convergence} shows the price RMSE on $G$ against run
time, for one feature draw of the network with $\nfeat$ from $300$ to
$3000$ features and for the implicit finite-difference scheme with $501$
to $8001$ grid points (and half as many time steps), each with and without
the subtraction. Run time covers assembling and solving the equations and
evaluating the prices on $G$, on one thread. Both solvers are written in
Python with NumPy and SciPy, so the comparison of run times is indicative
rather than definitive.

The plain network levels off between $5\times10^{-5}$ and $10^{-4}$ for
both models. This plateau is the effect of the payoff kink, which a finite
sum of smooth features cannot represent at short maturities
(Section~\ref{sec:bs-subtraction}). With the subtraction, the error falls
by a factor of about five as the features increase from $300$ to $1200$,
and then levels off near $7\times10^{-7}$ for Merton and $2\times10^{-6}$
for Kou. The ridge does not set this floor: changing it from $10^{-13}$
to $10^{-9}$ changes the Merton error by less than $25\%$. More collocation
points lower it only slowly; with twice the features and three times the
points the Merton error falls to $2.5\times10^{-7}$, for nine times the
run time. The finite-difference error, in contrast, keeps falling at
second order in the grid spacing.

As a result, the comparison depends on the run time. For short run times
the network with subtraction is the more accurate: for Merton, one draw
with $1200$ features reaches $8\times10^{-7}$ in $2.4$ seconds, which the
plain finite-difference scheme needs about $11$ seconds to match; for Kou
the network is slightly more accurate up to about eight seconds. Beyond
that the finite-difference scheme is the more accurate, and with the
subtraction it is more accurate than either at every run time, by about
two orders of magnitude. Averaging four feature draws, as in
Table~\ref{tab:jump-fdm}, lowers the network error by a factor of about
$1.5$ for four times the run time.}

\subsection{Calibration of the jump models}\label{sec:jump-calib}

This study shows that the exact Jacobian calibrates diffusion and jump
parameters alike, and that the solver's accuracy decides whether the
parameters are recovered, as Section~\ref{sec:identifiability} predicts.

The quotes are $36$ exact prices (the series for Merton,
COS for Kou) at the true parameters of Table~\ref{tab:setup}, for
$S/K\in\{0.80,0.85,\dots,1.20\}$ and $\tau\in\{0.25,0.5,0.75,1\}$. All four
Merton parameters and all five Kou parameters are calibrated from the
starting points in Table~\ref{tab:jump-calib}, \chgg{with one draw of
$800$ features, by the trust-region reflective method (Section~\ref{sec:calib-procedure})
with tolerances $10^{-12}$. The boundary conditions are those of
Section~\ref{sec:experiments}, so no reference prices are used.}

\begin{table}[htbp]
\centering
\caption{Calibration of the jump models to exact prices, with the plain
solve and with the Black--Scholes subtraction. \chgg{Recovered values are
for feature draw $0$; brackets give the range over five feature draws.
The pricing error $\varepsilon$ and the price RMSE are taken over the $36$
quotes. Bounds: $\sigma\in[0.01,1]$, $\lambda\in[0,3]$,
$\mu_J\in[-0.5,0.3]$, $\sigma_J\in[0.01,0.8]$, $p\in[0.05,0.95]$,
$\eta_1,\eta_2\in[1.5,30]$.}}
\label{tab:jump-calib}
\small
\begin{tabular}{llrrrrrr}
\toprule
Model & Parameter & True & Start & \multicolumn{2}{c}{Plain solve} & \multicolumn{2}{c}{With subtraction} \\
\midrule
\multirow{7}{*}{Merton}
 & $\sigma$   & $0.20$  & $0.30$ & \chgg{$0.1972$} & \chgg{$[0.189, 0.199]$} & \chgg{$0.2000$} & \chgg{$[0.1999, 0.2000]$} \\
 & $\lambda$  & $0.10$  & $0.30$ & \chgg{$0.1507$} & \chgg{$[0.105, 0.431]$} & \chgg{$0.1000$} & \chgg{$[0.0996, 0.1013]$} \\
 & $\mu_J$    & $-0.10$ & $0.00$ & \chgg{$-0.0771$} & \chgg{$[-0.102, -0.038]$} & \chgg{$-0.0999$} & \chgg{$[-0.1004, -0.0988]$} \\
 & $\sigma_J$ & $0.30$  & $0.20$ & \chgg{$0.2544$} & \chgg{$[0.167, 0.301]$} & \chgg{$0.3002$} & \chgg{$[0.2982, 0.3005]$} \\
 & \multicolumn{3}{l}{\chgg{pricing error $\varepsilon$ at the true parameters}} & \multicolumn{2}{c}{\chgg{$8.5\times10^{-5}$}} & \multicolumn{2}{c}{\chgg{$1.7\times10^{-6}$}} \\
 & \multicolumn{3}{l}{price RMSE, start}  & \multicolumn{2}{c}{\chgg{$0.0263$}} & \multicolumn{2}{c}{\chgg{$0.0262$}} \\
 & \multicolumn{3}{l}{price RMSE, final}  & \multicolumn{2}{c}{\chgg{$7.1\times10^{-5}$}} & \multicolumn{2}{c}{\chgg{$1.2\times10^{-6}$}} \\
\midrule
\multirow{8}{*}{Kou}
 & $\sigma$   & $0.20$  & $0.30$ & \chgg{$0.2016$} & \chgg{$[0.198, 0.202]$} & \chgg{$0.2001$} & \chgg{$[0.2000, 0.2001]$} \\
 & $\lambda$  & $0.30$  & $0.20$ & \chgg{$0.2354$} & \chgg{$[0.218, 0.412]$} & \chgg{$0.2937$} & \chgg{$[0.2937, 0.3033]$} \\
 & $p$        & $0.40$  & $0.55$ & \chgg{$0.2613$} & \chgg{$[0.204, 0.583]$} & \chgg{$0.3897$} & \chgg{$[0.3897, 0.4059]$} \\
 & $\eta_1$   & $10.00$ & $8.00$ & \chgg{$7.861$} & \chgg{$[6.84, 12.84]$} & \chgg{$9.817$} & \chgg{$[9.817, 10.11]$} \\
 & $\eta_2$   & $5.00$  & $6.50$ & \chgg{$4.938$} & \chgg{$[4.82, 5.36]$} & \chgg{$4.989$} & \chgg{$[4.989, 5.005]$} \\
 & \multicolumn{3}{l}{\chgg{pricing error $\varepsilon$ at the true parameters}} & \multicolumn{2}{c}{\chgg{$4.6\times10^{-5}$}} & \multicolumn{2}{c}{\chgg{$1.7\times10^{-6}$}} \\
 & \multicolumn{3}{l}{price RMSE, start}  & \multicolumn{2}{c}{\chgg{$0.0221$}} & \multicolumn{2}{c}{\chgg{$0.0220$}} \\
 & \multicolumn{3}{l}{price RMSE, final}  & \multicolumn{2}{c}{\chgg{$3.6\times10^{-5}$}} & \multicolumn{2}{c}{\chgg{$1.3\times10^{-6}$}} \\
\bottomrule
\end{tabular}
\end{table}

\begin{figure}[htbp]
\centering
\includegraphics[width=\textwidth]{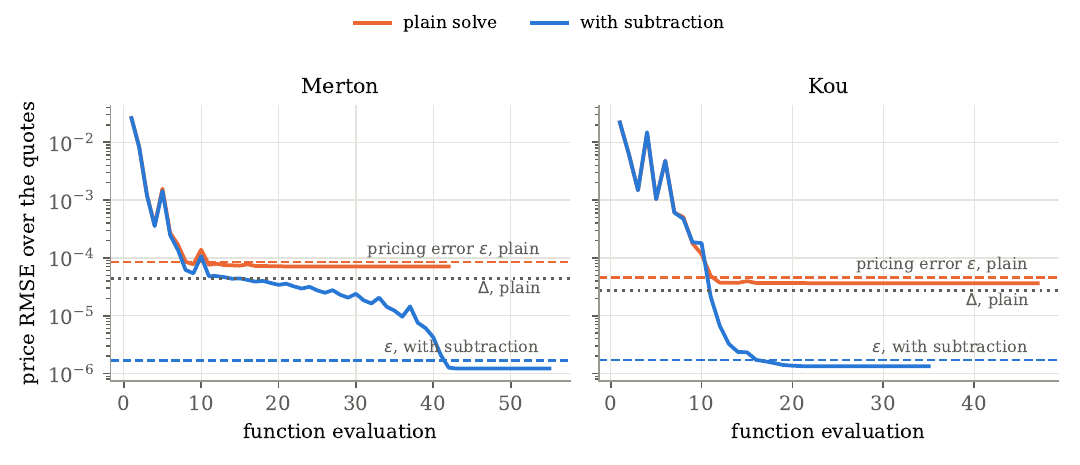}
\caption{\chgg{Calibration of the jump models: price RMSE over the quotes
at each function evaluation, for the plain solve and the solve with
subtraction (feature draw $0$). Dashed lines mark each solve's pricing
error $\varepsilon$ at the true parameters; the dotted line marks the
change $\Delta$ in the exact prices caused by the parameter error of the
plain solve.}}
\label{fig:kou-calib}
\end{figure}

\chgg{Both solves fit the quotes: the price RMSE falls by more than $99\%$
in every case (Table~\ref{tab:jump-calib}). Only the solve with
subtraction recovers the parameters, to within $0.1\%$ for Merton and
$3\%$ for Kou. With the plain solve the diffusion volatility is within
$1.5\%$, but the jump parameters are off by up to $51\%$ (Merton's jump
intensity $\lambda$) and $35\%$ (Kou's up-jump probability $p$).

The result does not depend on the feature draw or the starting point.
Over five feature draws, the solve with subtraction recovers every
parameter to within $3\%$ in every draw. The plain solve fits the quotes
in every draw too, but its jump parameters move with the draw: Merton's
$\lambda$ ranges from $0.11$ to $0.43$ and Kou's $p$ from $0.20$ to
$0.58$. Starting instead from $0.6$ and $1.4$ times the true values, both
solves return the same parameters as before, to within $0.5\%$, so the
plain solve's error is not a local minimum of the calibration.

The bound \eqref{eq:param-error} explains the difference. Let
$\varepsilon$ be a solve's pricing error at the true parameters, and
$\Delta$ the change in the exact prices between the true parameters and
those it returns, both as RMSE over the quotes. For the plain solve,
$\varepsilon=8.5\times10^{-5}$ and $\Delta=4.5\times10^{-5}$ for Merton,
and $\varepsilon=4.6\times10^{-5}$ and $\Delta=2.8\times10^{-5}$ for Kou.
The wrong parameters therefore change the exact prices by less than the
solver's own error, and the quotes cannot reveal them. With the
subtraction $\varepsilon=1.7\times10^{-6}$ for both models, $50$ and $27$
times smaller, and $\Delta$ ($1.2\times10^{-6}$ and $1.1\times10^{-6}$)
and the parameter errors shrink accordingly. With the smallest singular
value of the Jacobian in relative units ($3.1\times10^{-4}$ for Merton and
$1.1\times10^{-4}$ for Kou), the bound \eqref{eq:param-error} limits the
relative parameter error of the solve with subtraction to about $6\%$ and
$18\%$, and gives no useful limit for the plain solve. The actual errors
are well inside the bound, which allows for the worst direction in
parameter space. Figure~\ref{fig:kou-calib} shows the calibrations: the
plain solve stalls near its own pricing error, while the solve with
subtraction descends to its much lower one.

When the quotes are instead generated by the plain solve itself, with the
same feature draw, the plain solve recovers every parameter to within
$0.02\%$, with a final price RMSE of $1.4\times10^{-9}$ (Merton) and
$2.8\times10^{-9}$ (Kou). Testing a calibration on quotes produced by the
solver it uses is known in inverse problems as an \emph{inverse crime}
\citep{kaipio2007}: it hides the solver's error and proves little. Quotes
from the plain solve with a different feature draw behave like exact
quotes, with parameter errors of up to $43\%$ (Merton) and $40\%$ (Kou).
The loss of the parameters is therefore caused by the solver's error, not
by a lack of information in the quotes.

The repeated draws also show how to detect the problem without a
reference price, as Section~\ref{sec:identifiability} suggests. The
calibrated values of the plain solve scatter widely from draw to draw,
which marks them as unreliable, whereas those of the solve with
subtraction agree to within $3\%$.

Each Merton calibration needs $42$ function and $22$ Jacobian
evaluations with the plain solve ($55$ and $34$ with the subtraction) and
takes $34$ seconds ($46$ seconds); each Kou calibration needs $47$ and
$26$ ($35$ and $23$) and takes about four minutes (three minutes), on one
thread.}

\subsection{Comparison with a physics-informed neural network}\label{sec:bansal}

\citet{bansal2026} price Merton-type PIDEs with a PINN, approximating the
jump integral by the trapezoidal rule, on three manufactured test
problems. In Examples~1 and~2 the exact solution keeps a corner at the
strike for all times, so the manufactured right-hand side is itself
discontinuous there; Example~3, $\Vb=e^{x-t}$, is smooth.
Table~\ref{tab:bansal} compares their published results with ours.

On Examples~1 and~2 the plain solve is about as accurate as their PINN,
slightly less so on Example~1, and takes under three seconds instead of
about $1650$; their run times are as published, on different hardware.
\chgg{The subtraction of Section~\ref{sec:bs-subtraction}
works here too, but in a different form. A Black--Scholes price is smooth
for $\tau>0$ and cannot remove a corner that persists for all times, so
in Examples~1 and~2 we subtract instead the part of the manufactured
solution that carries the corner, and solve for the smooth remainder by
the same least-squares method, with $h$ in \eqref{eq:lifting} equal to
that part. The maximum error then falls below that of both by about two
orders of magnitude in Example~2 and three in Example~1. This function is available only because the exact solution is
known, and the PINN is not given it, so these rows show what the
subtraction can achieve when the kink is known exactly rather than a
like-for-like comparison.}
On the smooth Example~3, the Fourier features of Remark~\ref{sec:fourier}
reach a maximum error of $3\times10^{-8}$, several orders of magnitude
below the PINN and the probit features.

\begin{table}[htbp]
\centering
\caption{Comparison on the three test problems of \citet{bansal2026}.
PINN and FDM entries are as published; \PIRanNN{} entries are medians over
five feature draws on a $101\times101$ evaluation grid.}
\label{tab:bansal}
\adjustbox{max width=\textwidth}{%
\begin{tabular}{lllccc}
\toprule
Example & Method & Configuration & Max.\ error & MSE & CPU (s) \\
\midrule
\multirow{3}{*}{1 (corner, call-type)}
 & PINN \citep{bansal2026} & 300 epochs        & $4.95\times10^{-2}$ & $5.90\times10^{-5}$ & $1649$ \\
 & \PIRanNN{}, plain       & $\nfeat=800$      & $5.73\times10^{-2}$ & $8.64\times10^{-5}$ & $2.80$ \\
 & \PIRanNN{}, \chgg{known corner subtracted} &              & $1.77\times10^{-5}$ & $1.63\times10^{-11}$ & $0.77$ \\
\midrule
\multirow{3}{*}{2 (corner, put-type)}
 & PINN \citep{bansal2026} & 300 epochs        & $5.29\times10^{-2}$ & $6.51\times10^{-5}$ & $1672$ \\
 & \PIRanNN{}, plain       & $\nfeat=800$      & $5.69\times10^{-2}$ & $8.45\times10^{-5}$ & $2.28$ \\
 & \PIRanNN{}, \chgg{known corner subtracted} &              & $3.22\times10^{-4}$ & $4.54\times10^{-9}$ & $0.86$ \\
\midrule
\multirow{4}{*}{3 (smooth, $e^{x-t}$)}
 & FDM \citep{bansal2026}  & $512\times512$    & $5.83\times10^{-4}$ & $6.15\times10^{-8}$ & $454$ \\
 & PINN \citep{bansal2026} & 4 layers, 10 neurons & $1.48\times10^{-4}$ & $1.74\times10^{-9}$ & $348$ \\
 & \PIRanNN{}, probit      & $\nfeat=400$      & $7.27\times10^{-4}$ & $4.59\times10^{-8}$ & $0.37$ \\
 & \PIRanNN{}, Fourier     & $\nfeat=300$      & $3.01\times10^{-8}$ & $2.83\times10^{-17}$ & $1.49$ \\
\bottomrule
\end{tabular}}
\end{table}

\subsection{Heston}\label{sec:heston-exp}

This study shows that the method applies unchanged to an equation whose
coefficients depend on the state, with simple boundary conditions and no
condition at all on the degenerate face $v=0$.

\chgb{As boundary conditions, we} impose \eqref{eq:bc-asymptotic} on the two
faces in $x$ and nothing on the faces in $v$. \chg{At $v=0$ the diffusion
coefficients of \eqref{eq:heston-pde} vanish and, as is standard, no
condition is imposed: the equation itself determines the solution there
\citep{ekstrom2010}.} At $v_{\max}$ we also leave the solution to the
equation.

We price at the money, $S=K=100$ with $v_0=0.04$, in
the four test cases of \citet{inthout2010} and compare with three ADI
schemes from that study (Table~\ref{tab:heston-adi}).
The errors range from $0.02\%$ to $1.4\%$. The \chgo{standard error} over feature draws,
$0.4\%$ to $1.9\%$, is of the same size as the errors, so the differences
between the cases are largely within \chgo{it}. \chgb{The \chgo{variation between draws} is
therefore a useful indication of the error when no reference price is
available}\chgo{, as long as the error is not common to all draws (see
below)}\chgb{.} \PIRanNN{} is the most
accurate of the four methods in Case~3, which combines a Feller ratio
$2\kappa\vbar/\xi^2$ close to one with the longest maturity and is where
the ADI schemes show their largest errors. It is less accurate than the
MCS and HV schemes in Cases~1 and~2 and than all three schemes in
Case~4. Our domain in $v$ is also narrower than that of the ADI study.

\begin{figure}[htbp]
\centering
\includegraphics[width=\textwidth]{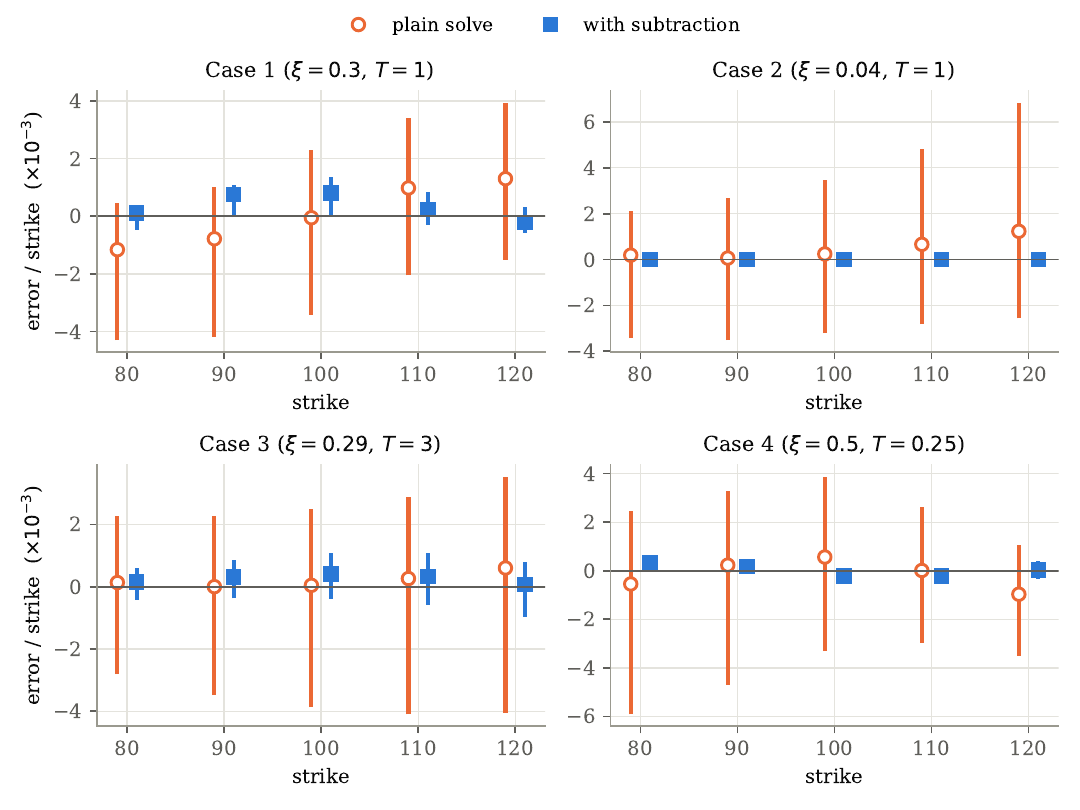}
\caption{\chgo{Heston: error of the call price at strikes $80$ to $120$
and the maturity of each case, relative to the strike, for the plain
solve and the solve with subtraction. Markers show the mean over the
eight feature draws and bars the range over the draws.}}
\label{fig:heston-strikes}
\end{figure}

\chgo{These results use the plain solve. To see what the subtraction of
Section~\ref{sec:bs-subtraction} adds, we repeat the four cases with it,
keeping the features, the collocation points and the boundary conditions;
only the right-hand side of the least-squares problem changes.
Table~\ref{tab:heston-sub} compares the two solves over the same eight
draws. Besides the at-the-money error, it gives the root-mean-square
error over the strikes $80$ to $120$ at the maturity of each case,
relative to the strike, computed for each draw and then averaged over the
draws. In every case the subtraction makes the solve much less dependent
on the feature draw: the standard error at the money falls by a factor of
between about $4$ and $250$, and the error over the strikes by a factor of
about $3$ in Cases~1 and~3, $6$ in Case~4 and $160$ in Case~2.
Figure~\ref{fig:heston-strikes} shows the error at each strike: the
spread over the draws shrinks at every strike. Relative to the price
rather than the strike, the gain is largest out of the money, where the
price is small and the plain errors reach tens of percent. In Case~2
the volatility of variance is very small, $\xi=0.04$, so the subtracted
price is already close to the Heston price.

At the money the picture is mixed. The typical error of a single draw is
smaller with the subtraction in every case: averaged over the draws, its
absolute value falls from $1.5\%$, $1.2\%$, $0.7\%$ and $4.2\%$ in Cases~1 to~4 to
$0.9\%$, $0.006\%$, $0.3\%$ and $0.5\%$. The mean error over the draws,
however, is not always smaller, because the plain errors take both signs
and partly cancel. With the subtraction, the error in Cases~1, 3 and~4 has
the same sign in most draws, with means of $+0.9\%$, $+0.2\%$ and $-0.5\%$.
Such an error is largely common to the draws, so it does not come from the
random features, and the spread over draws does not reveal it; it
presumably comes from a part of the set-up that all draws share, such as
the collocation grid or the boundary data, which we have not identified.
Against the ADI schemes of Table~\ref{tab:heston-adi}, the best of them
remains more accurate in Cases~1 and~4 and \PIRanNN{} in Case~3, with or
without the subtraction. In Case~2 the subtraction brings \PIRanNN{} to
about the accuracy of the MCS scheme, since its mean error is within one
standard error of zero, and ahead of the DO and HV schemes.}

\begin{table}[htbp]
\centering
\caption{Heston at-the-money prices: relative error of \PIRanNN{} (mean
over eight feature draws\chg{, each an average of three networks,} with the
standard error over draws) and of the
Douglas (DO), modified Craig--Sneyd (MCS) and Hundsdorfer--Verwer (HV)
ADI schemes reported by \citet{inthout2010}, against a Fourier reference.}
\label{tab:heston-adi}
\adjustbox{max width=\textwidth}{%
\begin{tabular}{c@{\hspace{2mm}}ccccccc@{\hspace{4mm}}rrrrr}
\toprule
Case & $\kappa$ & $\vbar$ & $\xi$ & $\rho$ & $r$ & $q$ & $T$ & Reference & \PIRanNN{} & DO & MCS & HV \\
\midrule
1 & $1.5$    & $0.04$   & $0.3$    & $-0.9$    & $0.025$  & $0$      & $1$    & $8.895$  & $-0.06\%$ ($0.70\%$) & $-0.412\%$ & $0.004\%$ & $-0.003\%$ \\
2 & $3$      & $0.12$   & $0.04$   & $0.6$     & $0.01$   & $0.04$   & $1$    & $10.542$ & $+0.23\%$ ($0.63\%$) & $-0.938\%$ & $0.000\%$ & $0.018\%$ \\
3 & $0.6067$ & $0.0707$ & $0.2928$ & $-0.7571$ & $0.03$   & $0$      & $3$    & $19.484$ & $+0.02\%$ ($0.36\%$) & $2.608\%$ & $3.188\%$ & $3.198\%$ \\
4 & $2.5$    & $0.06$   & $0.5$    & $-0.1$    & $0.0507$ & $0.0469$ & $0.25$ & $4.086$  & $+1.38\%$ ($1.85\%$) & $0.430\%$ & $0.977\%$ & $0.977\%$ \\
\bottomrule
\end{tabular}}
\end{table}

\begin{table}[htbp]
\centering
\caption{\chgo{Heston, plain solve and solve with subtraction of the price
with zero volatility of variance, over the same eight feature draws (the
plain values at the money are those of Table~\ref{tab:heston-adi}). At
the money: mean relative error, with the standard error over draws.
Strikes $80$ to $120$ at the maturity $T$ of each case: root-mean-square
error relative to the strike, computed for each draw and averaged over
the draws.}}
\label{tab:heston-sub}
\chgo{\begin{tabular}{ccc rr rr}
\toprule
 & & & \multicolumn{2}{c}{At the money} & \multicolumn{2}{c}{Strikes $80$ to $120$} \\
\cmidrule(lr){4-5}\cmidrule(lr){6-7}
Case & $\xi$ & $T$ & Plain & With subtraction & Plain & With subtraction \\
\midrule
1 & $0.3$    & $1$    & $-0.06\%$ ($0.70\%$) & $+0.91\%$ ($0.15\%$)   & $1.6\times10^{-3}$ & $5.9\times10^{-4}$ \\
2 & $0.04$   & $1$    & $+0.23\%$ ($0.63\%$) & $+0.003\%$ ($0.003\%$) & $1.6\times10^{-3}$ & $1.0\times10^{-5}$ \\
3 & $0.2928$ & $3$    & $+0.02\%$ ($0.36\%$) & $+0.21\%$ ($0.10\%$)   & $1.5\times10^{-3}$ & $5.3\times10^{-4}$ \\
4 & $0.5$    & $0.25$ & $+1.38\%$ ($1.85\%$) & $-0.52\%$ ($0.15\%$)   & $1.8\times10^{-3}$ & $2.7\times10^{-4}$ \\
\bottomrule
\end{tabular}}
\end{table}

We calibrate all five parameters
$(\kappa,\vbar,\xi,\rho,v_0)$ to $42$ Fourier prices at the Case~1
parameters, for six maturities from one month to one year and seven
strikes with $K/S\in[0.85,1.15]$. Table~\ref{tab:sv-calib} gives the
result. The current variance $v_0$, which only moves the evaluation point
(Section~\ref{sec:v0}), is recovered to $0.5\%$. Among the parameters of
the equation, $\vbar$ is recovered best ($2.9\%$) and $\kappa$ worst
($13.6\%$). \chgb{The calibration takes $19$ function and $18$ Jacobian
evaluations and about two minutes, and ends with a price RMSE of $0.023$
over the quotes (prices in units of a spot of $100$).

The Jacobian at the calibrated parameters shows why $\kappa$ is the
hardest parameter. First, $\kappa$ has little effect on these quotes: a
$1\%$ change in $\kappa$ moves them only about a sixth as much as a $1\%$
change in $\vbar$. Second, $\kappa$, $\xi$ and $\rho$ move the quotes in
almost the same way (their columns of the Jacobian are nearly parallel),
so an error in one can be largely compensated by the others, whereas
$\vbar$ acts differently and is the best-recovered parameter of the
equation. The parameter error therefore lies in directions that barely
change the quotes, which is the situation of
Section~\ref{sec:identifiability}. (The Jacobian agrees with finite
differences of the Fourier price to within $7\%$, apart from the
smallest $\kappa$ entries.)} The
correlation ends at its bound $\rho=-0.99$, so its value is not an
interior estimate. The fitted prices nevertheless match the quotes to a
mean relative error of $1\%$; in implied volatility, the error falls from
about $100$ basis points at one month to $4$ basis points at one year.

At the Case~1 parameters and strikes from $85$ to
$115$, Delta, Gamma and the sensitivity to $v_0$ agree with finite
differences of the Fourier price to within $2.3\%$, $5.9\%$ and $2.8\%$.
\chgb{Here the \PIRanNN{} values are also central differences, taken on
the fitted surface of one solve: the spot and $v_0$ are moved, but the
equation is not solved again.}

\subsection{SABR}\label{sec:sabr-exp}

This study shows that the method also handles coefficients that depend
nonlinearly on the state, with a boundary condition on only one of the
faces in $x$.

\chg{\chgb{As boundary conditions, we} impose the out-of-the-money condition
at $x_{\min}$ and, on the two faces in $\alpha$, the far-field condition
$\partial_{\alpha\alpha}\Vb=0$, which states that the price is
approximately linear in the volatility there.}

On a grid of five maturities from one month to two
years and five strikes with $K/F_0\in[0.7,1.3]$, restricted to the $18$
quotes with appreciable time value, \PIRanNN{} prices differ from Hagan's
formula by $0.5\%$ in root-mean-square and by at most $1.4\%$. \chg{Hagan's
formula is itself an approximation, so this is a measure of agreement, not
of error.} On a separate grid with maturities up to five years and strikes
up to $K/F_0=2$, compared with Monte Carlo simulation of
\eqref{eq:sabr-sde} \chg{($400{,}000$ paths, $500$ time steps)}, \PIRanNN{} is at least as close to the Monte Carlo
price as Hagan's formula at $9$ of the $10$ points. The exception is the
most extreme point, five years and $100\%$ out of the money, where the
errors are $28.6\%$ for \PIRanNN{} and $13.5\%$ for Hagan's formula.

\chgo{To test the subtraction of Section~\ref{sec:bs-subtraction}, we
need a reference more accurate than Hagan's formula. With $\beta=1$ and a
given path of $\alpha$, the logarithm of the forward at maturity is
normal, so the price is the average of Black prices over simulated paths
of $\alpha$ alone (conditional Monte Carlo; \citealp{glasserman2004}). We
simulate $\alpha$ exactly on a time grid, with $8$ million paths, and the
standard error is below $0.01\%$ of the price at every quote.
Table~\ref{tab:sabr-sub} gives the errors against this reference on the
$18$ quotes of the first grid. Hagan's formula has a root-mean-square
error of $0.21\%$ (at most $0.43\%$). For the price averaged over the
five feature draws, as above, the subtraction lowers the error of
\PIRanNN{} by a factor of about $10$, to about one fifth of the error of
Hagan's formula. It lowers the error of a single draw by a similar
factor, and with the subtraction every single draw is more accurate than
Hagan's formula.}

We calibrate $(\rho,\nu,\alpha_0)$ to \chg{$49$}
prices from Hagan's formula, for \chg{seven maturities from one month to
two years and seven strikes with $K/F_0\in[0.85,1.15]$}
(Table~\ref{tab:sv-calib}). The current volatility $\alpha_0$ is recovered
to $0.1\%$, the correlation to $10\%$ and the volatility of volatility to
$14\%$\chgb{, with a price RMSE of $0.023$ over the quotes, after $16$
function and $9$ Jacobian evaluations and $28$ seconds}. \chgb{The
situation differs from Heston: $\rho$ and $\nu$ move the quotes in clearly
different ways, so they do not compensate each other. Rather, the
remaining misfit is large compared with their effect on the quotes, and
the bound of Section~\ref{sec:identifiability} allows parameter errors of
up to about $20\%$.} \chg{Because the quotes come from Hagan's approximation rather
than from the SABR equation that the solver solves, part of these errors
reflects the difference between the two, and this test does not separate
it from the solver's error.} \chgo{On the pricing grid above, Hagan's
formula differs from the accurate reference by $0.21\%$ in
root-mean-square.}

At one year and strikes from $80$ to $120$, Delta and
the sensitivity to $\alpha_0$ agree with finite differences of Hagan's
formula to within $0.6\%$ and $1.6\%$\chgb{, again with central
differences on the fitted surface of one solve}.

\begin{table}[htbp]
\centering
\caption{\chgo{SABR: root-mean-square relative error on the $18$ quotes
of the pricing grid, against prices from conditional Monte Carlo
simulation. For \PIRanNN{}: the error of a single feature draw (mean
over the five draws, with the range in brackets), and the error of the
price averaged over the five draws. Hagan's formula involves no random
draws.}}
\label{tab:sabr-sub}
\chgo{\begin{tabular}{lcc}
\toprule
Method & Single draw & Price averaged over draws \\
\midrule
Hagan's formula & \multicolumn{2}{c}{$0.21\%$} \\
\PIRanNN{}, plain solve & $1.02\%$ ($0.60\%$ to $1.43\%$) & $0.47\%$ \\
\PIRanNN{}, with subtraction & $0.079\%$ ($0.036\%$ to $0.13\%$) & $0.044\%$ \\
\bottomrule
\end{tabular}}
\end{table}

\begin{table}[htbp]
\centering
\caption{Calibration of the stochastic-volatility models: Heston to
Fourier prices at the Case~1 parameters, SABR to prices from Hagan's
formula.}
\label{tab:sv-calib}
\begin{tabular}{llrrr}
\toprule
Model & Parameter & True & Recovered & Rel.\ error \\
\midrule
\multirow{5}{*}{Heston}
 & $\kappa$ & $1.500$  & $1.704$  & $13.6\%$ \\
 & $\vbar$  & $0.040$  & $0.0388$ & $2.9\%$ \\
 & $\xi$    & $0.300$  & $0.280$  & $6.8\%$ \\
 & $\rho$   & $-0.900$ & $-0.990$ & $10.0\%$ \\
 & $v_0$    & $0.040$  & $0.0402$ & $0.5\%$ \\
\midrule
\multirow{3}{*}{SABR}
 & $\rho$     & $-0.300$ & $-0.271$ & $9.8\%$ \\
 & $\nu$      & $0.400$  & $0.454$  & $13.6\%$ \\
 & $\alpha_0$ & $0.250$  & $0.250$  & $0.1\%$ \\
\bottomrule
\end{tabular}
\end{table}

\section{Conclusion}\label{sec:conclusion}

\chgo{We have presented a random-feature method for pricing and
calibrating options under jump-diffusion and stochastic-volatility
models. The output weights are found by one regularised least-squares
solve, the jump term of every feature is explicit for Merton and Kou, and
the same factorisation gives the Greeks and the exact Jacobian of the
prices with respect to the model parameters.

For the jump models, subtracting the Black--Scholes price removes the
error caused by the payoff kink. The method is then about as accurate as
a Crank--Nicolson scheme with implicit jump term at the same run time,
while a finite-difference scheme with the same subtraction is two orders
of magnitude more accurate. For Heston and SABR, boundary conditions that
need no reference prices are enough, and the plain solve prices to about
one percent; for Heston, the best ADI schemes are more accurate in three
of the four test cases. Subtracting the price with zero volatility of
volatility reduced the error for both models, by a factor of about $10$
for SABR and of $3$ to $160$ across the Heston cases, although for Heston
part of the error at the money remains. Solves take seconds and
calibrations minutes, far less than the published training times of a
PINN on the same test problems.

The calibrations show that fitting the quotes is not enough: the solver's
pricing error must be small compared with the effect of the parameters on
the quotes. Without the subtraction, the jump parameters are missed by up
to $50\%$; with it, they are recovered to within $3\%$. Repeating the
solve or the calibration with new feature draws reveals the part of the
error that varies between draws, but not errors common to all draws.

In one space dimension, if only prices are needed, a well-built
finite-difference scheme remains the better tool. The method is of
interest when one solve must also give a smooth price surface, exact
parameter sensitivities and a jump term without quadrature, in a
formulation that extends to more state variables.

The study is limited to European options in models with one or two
state variables and to synthetic, noise-free quotes. The loss weights
were chosen on these benchmarks, the method has no error bound, and its
accuracy levels off as features are added. Natural next steps are
barrier options, for which the jump term of a feature in the Merton
model remains explicit (a bivariate normal probability), Asian options,
exposure calculations along simulated paths,
other L\'evy models through Fourier features, and the use of the
subtraction in the calibration of Heston and SABR.}

\section*{Code availability}
\chgo{The code, with the settings used in the experiments and small
examples to try the method, is available at
\url{https://github.com/RiskLabIISc/PiRANN}.}

\appendix
\section{Parameter derivatives}\label{app:param-derivs}

This appendix lists the derivatives of the interior rows of $\mathbf A$
with respect to the model parameters, which enter
Proposition~\ref{prop:sens}. We write
$\calM_k=\calM[\Phi(s_k)]$ for the pricing operator applied to feature
$k$, evaluated at an interior point, and $a=A_k^{(x)}$. \chg{All formulas
were checked against finite differences.}

\chgb{For Merton, with} $\mathrm{sc}_k=\sqrt{1+a^2\sigma_J^2}$,
$u_k=(s_k+a\mu_J)/\mathrm{sc}_k$ and
$\kappa_J=e^{\mu_J+\sigma_J^2/2}-1$, so that the jump term of
\eqref{eq:M-merton-feature} is $-\lambda\Phi(u_k)$,
\begin{equation}\label{eq:merton-derivs}
	\left\{\begin{array}{l}
\partial_\sigma\calM_k = \sigma a\big(1+a s_k\big)\phi(s_k), \qquad
\partial_\lambda\calM_k = \kappa_J a\,\phi(s_k)+\Phi(s_k)-\Phi(u_k),\\ [8pt]
\displaystyle\partial_{\mu_J}\calM_k = \lambda a\Big((1+\kappa_J)\phi(s_k)-\frac{\phi(u_k)}{\mathrm{sc}_k}\Big), \qquad
\partial_{\sigma_J}\calM_k = \lambda\sigma_J\Big(a(1+\kappa_J)\phi(s_k)+\frac{a^2 u_k\,\phi(u_k)}{\mathrm{sc}_k^2}\Big).
\end{array}\right.
\end{equation}

\chgb{For Kou, write} the jump term as $-\lambda\mathcal N_k$ with
$\mathcal N_k=p\,G(s_k,a,\eta_1)+(1-p)\,G(s_k,-a,\eta_2)$ from
\eqref{eq:probit-kou-sum}, and
$\kappa_J=p\eta_1/(\eta_1-1)+(1-p)\eta_2/(\eta_2+1)-1$. Then
\begin{equation}\label{eq:kou-derivs}
\left\{\begin{array}{l}
\partial_\sigma\calM_k = \sigma a\,(1+a s_k)\,\phi(s_k), \qquad
\partial_\lambda\calM_k = \kappa_J a\,\phi(s_k)+\Phi(s_k)-\mathcal N_k,\\ [8pt]
\partial_p\calM_k = \lambda a\,\partial_p\kappa_J\,\phi(s_k)
   -\lambda\big(G(s_k,a,\eta_1)-G(s_k,-a,\eta_2)\big),\\ [8pt]
\partial_{\eta_i}\calM_k = \lambda a\,\partial_{\eta_i}\kappa_J\,\phi(s_k)
   -\lambda\,\partial_{\eta_i}\mathcal N_k ,
\end{array}\right.
\end{equation}
with $\partial_p\kappa_J=\eta_1/(\eta_1-1)-\eta_2/(\eta_2+1)$,
$\partial_{\eta_1}\kappa_J=-p/(\eta_1-1)^2$,
$\partial_{\eta_2}\kappa_J=(1-p)/(\eta_2+1)^2$,
$\partial_{\eta_1}\mathcal N_k=p\,\partial_\eta G(s_k,a,\eta_1)$,
$\partial_{\eta_2}\mathcal N_k=(1-p)\,\partial_\eta G(s_k,-a,\eta_2)$, and
\[
\partial_\eta G(s,a,\eta)=\frac{\mathrm{sgn}(a)}{2\sqrt2\,|a|}\,e^{-s^2/2}\,
\operatorname{erfcx}'\!\Big(\frac{as+\eta}{\sqrt2\,|a|}\Big),
\qquad
\operatorname{erfcx}'(w)=2w\operatorname{erfcx}(w)-\frac{2}{\sqrt\pi}.
\]

\chgb{For Heston, it follows from} \eqref{eq:PkQk-heston} \chgb{that}
\begin{equation}\label{eq:heston-derivs}
\left\{\begin{array}{l}
\partial_\kappa\calM_k = -(\vbar-v)\,A_k^{(v)}\phi(s_k), \qquad
\partial_{\vbar}\calM_k = -\kappa A_k^{(v)}\phi(s_k),\\ [6pt]
\displaystyle \partial_\xi\calM_k = \big(\rho v\,a A_k^{(v)}+\xi v\big(A_k^{(v)}\big)^2\big)s_k\,\phi(s_k), \qquad
\partial_\rho\calM_k = \xi v\,a A_k^{(v)}s_k\,\phi(s_k).
\end{array}\right.
\end{equation}
The sensitivity to $v_0$ is given by \eqref{eq:v0-sens}.

\chgb{For SABR,} \chg{it follows from \eqref{eq:PkQk-sabr} that the
two operator parameters enter only through $P_k$, and
\begin{equation}\label{eq:sabr-derivs}
 \partial_\rho\calM_k = \nu\alpha^{2} F^{\beta}A_k^{(F)}A_k^{(\alpha)}\,s_k\phi(s_k),
\qquad
\partial_\nu\calM_k = \big(\rho\alpha^{2} F^{\beta}A_k^{(F)}A_k^{(\alpha)}
  +\nu\alpha^2\big(A_k^{(\alpha)}\big)^2\big)\,s_k\phi(s_k).
\end{equation}}
The sensitivity to $\alpha_0$ is the state derivative of
Section~\ref{sec:v0}.

When the Black--Scholes price is
subtracted, the interior targets are the values of $\calF$ in
\eqref{eq:bs-source}. Its parameter derivatives follow by differentiating
\eqref{eq:JuBS-merton} or \eqref{eq:JuBS-kou} and the Black--Scholes price
and delta, all of which are explicit.

\section{Robustness to the jump parameters}\label{app:robustness}

Table~\ref{tab:mert-robust} repeats the Merton pricing study of
Section~\ref{sec:merton-exp} while varying one jump parameter at a time,
with all solver settings unchanged. \chgg{These runs were made with an
earlier version of the solver, which took its boundary values from the
reference price and measured errors on a coarser grid, so their absolute
errors are not comparable with those of Section~\ref{sec:merton-exp}; what
they show is the gain from the subtraction at each parameter set.} The plain solve stays near $10^{-4}$
throughout, and the Black--Scholes subtraction is about two orders of
magnitude more accurate for every parameter set.

\begin{table}[htbp]
\centering
\caption{Merton price RMSE when one jump parameter is varied, with the
solver settings unchanged. Bold marks the value used in the rest of the
paper.}
\label{tab:mert-robust}
\begin{tabular}{llcc}
\toprule
Parameter & Value & Plain solve & With subtraction \\
\midrule
\multirow{4}{*}{Intensity $\lambda$}
 & $0.05$ & $8.62\times10^{-5}$ & $2.64\times10^{-7}$ \\
 & $\mathbf{0.10}$ & $8.88\times10^{-5}$ & $5.37\times10^{-7}$ \\
 & $0.30$ & $9.75\times10^{-5}$ & $1.71\times10^{-6}$ \\
 & $0.50$ & $9.96\times10^{-5}$ & $2.93\times10^{-6}$ \\
\midrule
\multirow{4}{*}{Jump volatility $\sigma_J$}
 & $0.10$ & $8.50\times10^{-5}$ & $8.21\times10^{-7}$ \\
 & $0.20$ & $8.49\times10^{-5}$ & $7.06\times10^{-7}$ \\
 & $\mathbf{0.30}$ & $8.88\times10^{-5}$ & $5.37\times10^{-7}$ \\
 & $0.50$ & $3.96\times10^{-4}$ & $1.63\times10^{-6}$ \\
\midrule
\multirow{4}{*}{Jump mean $\mu_J$}
 & $-0.30$ & $8.72\times10^{-5}$ & $2.31\times10^{-6}$ \\
 & $\mathbf{-0.10}$ & $8.88\times10^{-5}$ & $5.37\times10^{-7}$ \\
 & $0.00$ & $9.85\times10^{-5}$ & $5.44\times10^{-7}$ \\
 & $0.10$ & $1.16\times10^{-4}$ & $1.75\times10^{-6}$ \\
\bottomrule
\end{tabular}
\end{table}

\bibliographystyle{plainnat}
\bibliography{references}

@article{bansal2025,
  author  = {Bansal, S. and Boro, P. and Natesan, S.},
  title   = {Physics-informed neural network for option pricing weather derivatives model},
  journal = {Computers \& Mathematics with Applications},
  volume  = {200}, pages = {1--21}, year = {2025},
  doi     = {10.1016/j.camwa.2025.09.001}
}

@article{bansal2026,
  author  = {Bansal, S. and Boro, P. and Natesan, S.},
  title   = {Application of physics informed neural networks to partial integro-differential equations in financial modeling and decision making},
  journal = {Applied Soft Computing},
  volume  = {186}, pages = {114208}, year = {2026}
}

@article{beck2021,
  author  = {Beck, C. and Becker, S. and Cheridito, P. and Jentzen, A. and Neufeld, A.},
  title   = {Deep splitting method for parabolic {PDE}s},
  journal = {SIAM Journal on Scientific Computing},
  volume  = {43}, number = {5}, pages = {A3135--A3154}, year = {2021}
}

@article{carr1999,
  author  = {Carr, P. and Madan, D. B.},
  title   = {Option valuation using the fast {F}ourier transform},
  journal = {Journal of Computational Finance},
  volume  = {2}, number = {4}, pages = {61--73}, year = {1999}
}

@article{chen2021,
  author  = {Chen, Y. and Yu, H. and Meng, X. and Xie, X. and Hou, M. and Chevallier, J.},
  title   = {Numerical solving of the generalized {B}lack--{S}choles differential equation using {L}aguerre neural network},
  journal = {Digital Signal Processing},
  volume  = {112}, pages = {103003}, year = {2021},
  doi     = {10.1016/j.dsp.2021.103003}
}

@article{chen2022rfm,
  author  = {Chen, J. and Chi, X. and E, W. and Yang, Z.},
  title   = {Bridging traditional and machine learning-based algorithms for solving {PDE}s: the random feature method},
  journal = {Journal of Machine Learning},
  volume  = {1}, number = {3}, pages = {268--298}, year = {2022}
}

@article{cont2005,
  author  = {Cont, R. and Voltchkova, E.},
  title   = {A finite difference scheme for option pricing in jump diffusion and exponential {L}\'evy models},
  journal = {SIAM Journal on Numerical Analysis},
  volume  = {43}, number = {4}, pages = {1596--1626}, year = {2005}
}

@article{dong2021,
  author  = {Dong, S. and Li, Z.},
  title   = {Local extreme learning machines and domain decomposition for solving linear and nonlinear partial differential equations},
  journal = {Computer Methods in Applied Mechanics and Engineering},
  volume  = {387}, pages = {114129}, year = {2021}
}

@article{dwivedi2020,
  author  = {Dwivedi, V. and Srinivasan, B.},
  title   = {Physics informed extreme learning machine ({PIELM}) -- a rapid method for the numerical solution of partial differential equations},
  journal = {Neurocomputing},
  volume  = {391}, pages = {96--118}, year = {2020}
}

@article{fang2008,
  author  = {Fang, F. and Oosterlee, C. W.},
  title   = {A novel pricing method for {E}uropean options based on {F}ourier-cosine series expansions},
  journal = {SIAM Journal on Scientific Computing},
  volume  = {31}, number = {2}, pages = {826--848}, year = {2008}
}

@article{fix1973,
  author  = {Fix, G. J. and Gulati, S. and Wakoff, G. I.},
  title   = {On the use of singular functions with finite element approximations},
  journal = {Journal of Computational Physics},
  volume  = {13}, number = {2}, pages = {209--228}, year = {1973}
}

@article{golub1973,
  author  = {Golub, G. H. and Pereyra, V.},
  title   = {The differentiation of pseudo-inverses and nonlinear least squares problems whose variables separate},
  journal = {SIAM Journal on Numerical Analysis},
  volume  = {10}, number = {2}, pages = {413--432}, year = {1973}
}

@article{gonon2023,
  author  = {Gonon, L.},
  title   = {Random feature neural networks learn {B}lack--{S}choles type {PDE}s without curse of dimensionality},
  journal = {Journal of Machine Learning Research},
  volume  = {24}, number = {189}, pages = {1--51}, year = {2023}
}

@article{hagan2002,
  author  = {Hagan, P. S. and Kumar, D. and Lesniewski, A. S. and Woodward, D. E.},
  title   = {Managing smile risk},
  journal = {Wilmott Magazine},
  pages   = {84--108}, year = {2002}
}

@article{hainaut2024,
  author  = {Hainaut, D. and Casas, A.},
  title   = {Option pricing in the {H}eston model with physics inspired neural networks},
  journal = {Annals of Finance},
  volume  = {20}, number = {3}, pages = {353--376}, year = {2024},
  doi     = {10.1007/s10436-024-00452-7}
}

@article{han2018,
  author  = {Han, J. and Jentzen, A. and E, W.},
  title   = {Solving high-dimensional partial differential equations using deep learning},
  journal = {Proceedings of the National Academy of Sciences},
  volume  = {115}, number = {34}, pages = {8505--8510}, year = {2018}
}

@article{heston1993,
  author  = {Heston, S. L.},
  title   = {A closed-form solution for options with stochastic volatility with applications to bond and currency options},
  journal = {Review of Financial Studies},
  volume  = {6}, number = {2}, pages = {327--343}, year = {1993}
}

@article{inthout2010,
  author  = {In 't Hout, K. J. and Foulon, S.},
  title   = {{ADI} finite difference schemes for option pricing in the {H}eston model with correlation},
  journal = {International Journal of Numerical Analysis and Modeling},
  volume  = {7}, number = {2}, pages = {303--320}, year = {2010}
}

@misc{jacquier2023,
  author  = {Jacquier, A. and Zuric, Z.},
  title   = {Random neural networks for rough volatility},
  year    = {2023}, howpublished = {arXiv:2305.01035}
}

@article{kou2002,
  author  = {Kou, S. G.},
  title   = {A jump-diffusion model for option pricing},
  journal = {Management Science},
  volume  = {48}, number = {8}, pages = {1086--1101}, year = {2002}
}

@article{merton1976,
  author  = {Merton, R. C.},
  title   = {Option pricing when underlying stock returns are discontinuous},
  journal = {Journal of Financial Economics},
  volume  = {3}, number = {1--2}, pages = {125--144}, year = {1976}
}

@techreport{mrad2025,
  author      = {Mrad, M. and Iassamen, N.},
  title       = {Differential deep learning and physics-informed neural networks for fast derivative pricing},
  institution = {HAL open archive},
  type        = {Working paper},
  number      = {hal-05160171}, year = {2025}
}

@article{neufeld2025,
  author  = {Neufeld, A. and Schmocker, P. and Wu, S.},
  title   = {Full error analysis of the random deep splitting method for nonlinear parabolic {PDE}s and {PIDE}s},
  journal = {Communications in Nonlinear Science and Numerical Simulation},
  volume  = {143}, pages = {108556}, year = {2025}
}

@misc{neufeldschmocker2023,
  author  = {Neufeld, A. and Schmocker, P.},
  title   = {Universal approximation property of {B}anach space-valued random feature models including random neural networks},
  year    = {2023}, howpublished = {arXiv:2312.08410}
}

@article{pooley2003,
  author  = {Pooley, D. M. and Vetzal, K. R. and Forsyth, P. A.},
  title   = {Convergence remedies for non-smooth payoffs in option pricing},
  journal = {Journal of Computational Finance},
  volume  = {6}, number = {4}, pages = {25--40}, year = {2003}
}

@inproceedings{rahimi2008,
  author    = {Rahimi, A. and Recht, B.},
  title     = {Weighted sums of random kitchen sinks: Replacing minimization with randomization in learning},
  booktitle = {Advances in Neural Information Processing Systems},
  volume    = {21}, pages = {1313--1320}, year = {2008}
}

@article{raissi2019,
  author  = {Raissi, M. and Perdikaris, P. and Karniadakis, G. E.},
  title   = {Physics-informed neural networks: A deep learning framework for solving forward and inverse problems involving nonlinear partial differential equations},
  journal = {Journal of Computational Physics},
  volume  = {378}, pages = {686--707}, year = {2019}
}

@inproceedings{sitzmann2020,
  author    = {Sitzmann, V. and Martel, J. N. P. and Bergman, A. W. and Lindell, D. B. and Wetzstein, G.},
  title     = {Implicit neural representations with periodic activation functions},
  booktitle = {Advances in Neural Information Processing Systems},
  volume    = {33}, pages = {7462--7473}, year = {2020}
}

@misc{srinivasan2025,
  author  = {Srinivasan, A. G. and Said, A. J. and Pentela, S. and Dwivedi, V. and Srinivasan, B.},
  title   = {Towards fast option pricing {PDE} solvers powered by {PIELM}},
  year    = {2025}, howpublished = {arXiv:2510.04322}
}

@article{vonsydow2019,
  author  = {von Sydow, L. and Milovanovi{\'c}, S. and Larsson, E. and In 't Hout, K. and Wiktorsson, M. and Oosterlee, C. W. and Shcherbakov, V. and Wyns, M. and Leitao, A. and Jain, S. and Haentjens, T. and Wald{\'e}n, J.},
  title   = {{BENCHOP--SLV}: the {BENCH}marking project in {O}ption {P}ricing -- {S}tochastic and {L}ocal {V}olatility problems},
  journal = {International Journal of Computer Mathematics},
  volume  = {96}, number = {10}, pages = {1910--1923}, year = {2019},
  doi     = {10.1080/00207160.2018.1544368}
}

@article{wang2023,
  author  = {Wang, Y. and Dong, S.},
  title   = {An extreme learning machine-based method for computational {PDE}s in higher dimensions},
  journal = {Computer Methods in Applied Mechanics and Engineering},
  volume  = {418}, pages = {116578}, year = {2024}
}

@article{albani2020,
  author  = {Albani, V. and Zubelli, J. P.},
  title   = {A splitting strategy for the calibration of jump-diffusion models},
  journal = {Finance and Stochastics},
  volume  = {24}, number = {3}, pages = {677--722}, year = {2020},
  doi     = {10.1007/s00780-020-00425-4}
}

@article{bae2024,
  author  = {Bae, H.-O. and Kang, S. and Lee, M.},
  title   = {Option pricing and local volatility surface by physics-informed neural network},
  journal = {Computational Economics},
  volume  = {64}, number = {5}, pages = {3143--3159}, year = {2024}
}

@misc{chakhoyan2025,
  author  = {Chakhoyan, H. and Kalici, S. and Malandain, K. A.},
  title   = {{DeepSVM}: Learning stochastic volatility models with physics-informed deep operator networks},
  year    = {2025}, howpublished = {arXiv:2512.07162}
}

@article{ipinn2026,
  author  = {Raeisi-Makiani, M. and Neisy, A. and Safdari-Vaighani, A.},
  title   = {The application of the inverse physics-informed neural network in financial calibration tasks},
  journal = {Journal of Mathematics and Modeling in Finance},
  volume  = {6}, number = {2}, pages = {127--139}, year = {2026}
}

@article{ekstrom2010,
  author  = {Ekstr{\"o}m, E. and Tysk, J.},
  title   = {The {B}lack--{S}choles equation in stochastic volatility models},
  journal = {Journal of Mathematical Analysis and Applications},
  volume  = {368}, number = {2}, pages = {498--507}, year = {2010}
}

@article{owen1980,
  author  = {Owen, D. B.},
  title   = {A table of normal integrals},
  journal = {Communications in Statistics -- Simulation and Computation},
  volume  = {9}, number = {4}, pages = {389--419}, year = {1980}
}

@book{shreve2004,
  author    = {Shreve, S. E.},
  title     = {Stochastic Calculus for Finance {II}: Continuous-Time Models},
  publisher = {Springer}, address = {New York}, year = {2004}
}

@book{cont2004book,
  author    = {Cont, R. and Tankov, P.},
  title     = {Financial Modelling with Jump Processes},
  publisher = {Chapman \& Hall/CRC}, address = {Boca Raton, FL}, year = {2004}
}

@article{branch1999,
  author  = {Branch, M. A. and Coleman, T. F. and Li, Y.},
  title   = {A subspace, interior, and conjugate gradient method for large-scale bound-constrained minimization problems},
  journal = {SIAM Journal on Scientific Computing},
  volume  = {21}, number = {1}, pages = {1--23}, year = {1999}
}

@article{sirignano2018,
  author  = {Sirignano, J. and Spiliopoulos, K.},
  title   = {{DGM}: A deep learning algorithm for solving partial differential equations},
  journal = {Journal of Computational Physics},
  volume  = {375}, pages = {1339--1364}, year = {2018}
}

@article{horvath2021,
  author  = {Horvath, B. and Muguruza, A. and Tomas, M.},
  title   = {Deep learning volatility: a deep neural network perspective on pricing and calibration in (rough) volatility models},
  journal = {Quantitative Finance},
  volume  = {21}, number = {1}, pages = {11--27}, year = {2021}
}

@article{dhalluin2005,
  author  = {d'Halluin, Y. and Forsyth, P. A. and Vetzal, K. R.},
  title   = {Robust numerical methods for contingent claims under jump diffusion processes},
  journal = {IMA Journal of Numerical Analysis},
  volume  = {25},
  number  = {1},
  pages   = {87--112},
  year    = {2005}
}

@article{kaipio2007,
  author  = {Kaipio, J. and Somersalo, E.},
  title   = {Statistical inverse problems: discretization, model reduction and inverse crimes},
  journal = {Journal of Computational and Applied Mathematics},
  volume  = {198},
  number  = {2},
  pages   = {493--504},
  year    = {2007}
}

@article{grossmann2024,
  author  = {Grossmann, T. G. and Komorowska, U. J. and Latz, J. and Sch{\"o}nlieb, C.-B.},
  title   = {Can physics-informed neural networks beat the finite element method?},
  journal = {IMA Journal of Applied Mathematics},
  volume  = {89}, number = {1}, pages = {143--174}, year = {2024},
  doi     = {10.1093/imamat/hxae011}
}

@article{mcgreivy2024,
  author  = {McGreivy, N. and Hakim, A.},
  title   = {Weak baselines and reporting biases lead to overoptimism in machine learning for fluid-related partial differential equations},
  journal = {Nature Machine Intelligence},
  volume  = {6}, number = {10}, pages = {1256--1269}, year = {2024},
  doi     = {10.1038/s42256-024-00897-5}
}

@book{lewis2000,
  author    = {Lewis, Alan L.},
  title     = {Option Valuation under Stochastic Volatility},
  publisher = {Finance Press},
  address   = {Newport Beach, CA},
  year      = {2000}
}

@article{benhamou2010,
  author  = {Benhamou, Eric and Gobet, Emmanuel and Miri, Mohammed},
  title   = {Time dependent {H}eston model},
  journal = {SIAM Journal on Financial Mathematics},
  volume  = {1},
  number  = {1},
  pages   = {289--325},
  year    = {2010},
  doi     = {10.1137/090753814}
}

@book{glasserman2004,
  author    = {Glasserman, Paul},
  title     = {Monte {C}arlo Methods in Financial Engineering},
  publisher = {Springer},
  address   = {New York},
  year      = {2004}
}

\end{document}